\documentclass[twocolumn,8pt]{extarticle}
\usepackage[font=small,labelfont=bf]{caption} 
\usepackage{graphicx}
\usepackage{authblk}
\usepackage{amsmath}
\usepackage{multicol}
\usepackage[square]{natbib}
\newcommand*\inst[1]{\unskip\hbox{\@textsuperscript{\normalfont$#1$}}}
\newcommand*\fnmsep{\unskip\hbox{\@textsuperscript{\normalfont,}}}
\def\ms{\ifmmode{\rm m\thinspace s^{-1}}\else m\thinspace s$^{-1}$\fi}
\def\degree{\ifmmode{^{\circ}}\else $^{\circ}$\fi}

\newcommand*\tablefootmark[1]{%
  \unskip
  \hbox{\@textsuperscript{\normalfont\itshape\ignorespaces#1}}%
  \,%
  \ignorespaces
}
\newcommand\tablefoottext[2]{%
  \hbox{\@textsuperscript{\normalfont({\itshape\ignorespaces#1})}}%
  ~%
  \ignorespaces
  #2\ \ignorespaces%
}
\newcommand\tablefoot[1]{\VSpaceBeforeTabBib=1ex%
  \par\vspace{\VSpaceBeforeTabFoot}
  \noindent
  \begin{minipage}{\linewidth}
    {\aa@tablefootnamefont\aa@tablefootname.}~%
    \aa@tablefootfont
    \ignorespaces
    #1%
  \end{minipage}%
}

\begin{document}

   \title{\vspace{-2.0cm}\Huge{\textbf{Efficient scheduling of astronomical observations}} \\
   \textbf{\LARGE{Application to the CARMENES radial-velocity survey}}}

   \author[1]{A.~Garcia-Piquer}
   \author[1,2]{J.~C.~Morales}
   \author[1]{I.~Ribas}
   \author[1]{J.~Colom\'e}
   \author[1]{J.~Gu\`ardia}
   \author[1]{M.~Perger}
   \author[3,4]{J.~A.~Caballero}
   \author[5]{M.~Cort\'es-Contreras}
   \author[6]{S.~V.~Jeffers}
   \author[6]{A.~Reiners}
   \author[7]{P.~J.~Amado}
   \author[3]{A.~Quirrenbach}
   \author[3]{W.~Seifert}
  
  \affil[1]{\small Institut de Ci\`encies de l'Espai (IEEC-CSIC), Campus UAB, C/Can Magrans s/n, 08193 Bellaterra, Spain, e-mail: agarcia@ice.csic.es}
  \affil[2]{\small {LESIA-Observatoire~de~Paris,~CNRS,~UPMC~Univ.~Paris~06,~Univ.~Paris-Diderot,~5~Pl.~Jules~Janssen,~92195~Meudon~CEDEX,~France}}
  \affil[3]{\small Landessternwarte, Zentrum f\"ur Astronomie der Universit\"at Heidelberg, K\"onigstuhl 12, 69117 Heidelberg, Germany}
  \affil[4]{\small Centro de Astrobiolog\'ia (CSIC-INTA), Camino Bajo del Castillo, 28691 Villanueva de la Ca\~nada, Madrid, Spain}
  \affil[5]{\small {Departamento~de~Astrof\'isica~y~Ciencias~de~la~Atm\'osfera,~Facultad~de~Ciencias~F\'isicas,~Universidad~Complutense,~28040~Madrid,~Spain}}
  \affil[6]{\small Institut f\"ur Astrophysik, Friedrich-Hund-Platz 1, 37077 G\"ottingen, Germany}
  \affil[7]{\small Instituto de Astrof\'isica de Andaluc\'ia (CSIC), Glorieta de la Astronom\'ia s/n, 18008 Granada, Spain}
%   \date{Received 11 March 2016}

% \abstract{}{}{}{}{}
% 5 {} token are mandatory
 \twocolumn[
  \begin{@twocolumnfalse}
  \maketitle
  \abstract
  % {Context heading (optional)}
   {Targeted spectroscopic exoplanet surveys face the challenge of maximizing
their planet detection rates by means of careful planning. The number of
possible observation combinations for a large exoplanet survey, i.e., the
sequence of observations night after night, both in total time and amount of
targets, is enormous.}
    {Sophisticated scheduling tools and the improved understanding of the
exoplanet population are employed to investigate an efficient and optimal way
to plan the execution of observations. This is applied to the CARMENES
instrument, which is an optical and infrared high-resolution spectrograph that
has started a survey of about 300 M-dwarf stars in search for terrestrial
exoplanets.}
   {We use evolutionary computation techniques to create an automatic scheduler
that minimizes the idle periods of the telescope and that distributes the
observations among all the targets using configurable criteria. We simulate the
case of the CARMENES survey with a realistic sample of targets, and we estimate
the efficiency of the planning tool both in terms of telescope operations and
planet detection.}
   {Our scheduling simulations produce plans that use about 99\% of the
available telescope time (including overheads) and optimally distribute the
observations among the different targets. Under such conditions, and using
current planet statistics, the optimized plan using this tool should allow
the CARMENES survey to discover about 65\% of the planets with radial-velocity
semi-amplitudes greater than 1~\ms\/ when considering only photon
noise.}
   {The simulations using our scheduling tool show that it is possible to
optimize the survey planning by minimizing idle instrument periods and
fulfilling the science objectives in an efficient manner to maximze the
scientific return.\\}
  \end{@twocolumnfalse}
 ]
 
%
%________________________________________________________________

\section{Introduction}
\label{sec:introduction}
Radial-velocity surveys, together with
dedicated photometric space missions such as \emph{CoRoT} and \emph{Kepler},
have proved to be the most efficient way of discovering exoplanets. Radial-velocity
surveys have generally been focused on solar-type stars, but recently
the interest in planets orbiting late-type stars has increased. Due to their
lower mass, the radial-velocity semi-amplitude induced by rocky planets around
M dwarfs is 1.5$-$5 times larger than in the case of solar-type stars for
similar orbital periods. This makes low-mass stars uniquely suited to the
detection of Earth-like planets provided that instruments reaching the 1~\ms\
precision level are available. Particularly interesting are rocky planets in
the habitable zones of their host stars, whose radial-velocity signals can
reach a few \ms\ in the case of M dwarfs. However, a spectroscopic survey of
low-mass stars has to face two main challenges: \emph{1)} the targets are
generally faint and \emph{2)} they show higher levels of intrinsic variability
due to higher stellar magnetic activity than Sun-like stars. For this reason, less
than 100 exoplanets out of the over 3500 known to date have been discovered
orbiting M dwarfs
\citep[e.g.,][]{Bonfils13b,Delfosse13,Tuomi14,AstudilloDefru15}, and most
of them are around early-M dwarfs.

CARMENES\footnote{{http://carmenes.caha.es}} (\textit{Calar Alto
high-Resolution search for M dwarfs with Exo-earths with Near-infrared and
optical \'Echelle Spectrographs}) is a next-generation instrument aiming at the
discovery and study of a statistically significant sample of exoplanets around
M dwarfs using precise radial velocities. CARMENES is mounted on the 3.5-m
Zeiss telescope at the Calar Alto Observatory \citep[Almer\'ia,
Spain;][]{Sanchez07,Sanchez08}, and it has been built by a German-Spanish
consortium \citep{Quirrenbach14}. It consists of two spectrographs, one
sensitive to visible light and another one to the near infrared. The radial
velocity precision is expected to be of the order of 1~\ms, similar to current
instruments in the visible, such as HARPS \citep{Mayor03} and HARPS-North
\citep{Cosentino12}. However, the main advantage of CARMENES is the
simultaneous measurement of Doppler shifts over a very wide spectral range
from the visible ($>$0.52~$\mu$m) to the near infrared ($<$1.71~$\mu$m). The
infrared channel is designed to monitor the radial velocity of stars in the
wavelength region of the spectra where late-type dwarfs emit the bulk of their
light. On the other hand, the visible channel has as well plenty of radial
velocity information (spectral lines are abundant), and the combination with
infrared measurements is ideal to monitor stellar activity, thus providing
means of disentangling true exoplanet signals from other effects
\citep{Reiners10,Reiners13}. As part of the guaranteed-time observations
(GTO), CARMENES is surveying about 300 M dwarfs \citep{Caballero13a,AlonsoFloriano15}.
The commissioning of the instrument ended in December 2015 and regular operations
started in January 2016. The GTO survey will extend for at least three years,
using over 600 nights of telescope time, with the goal of discovering dozens of
new planets, particularly focusing on those residing in the habitable zones of
their stars.

One of the major challenges of the CARMENES survey is the efficient planning
of the observations of the numerous targets in the sample. In general,
any kind of astronomical survey requires the execution of a large
number of observations fulfilling several constraints. Some of these
constraints can be predicted (e.g., visibility and elevation of the object) and
have to be necessarily satisfied, and others are unknown until the time of
execution of the observations (e.g., integration time, environmental
conditions). In addition, there are some scientific constraints that should be
optimized, such as the number of targets that have to be observed and the
number of observations of each target \citep{Perger2017}. The optimization of these constraints is
a key factor for obtaining a suitable schedule with an adequate exploitation of
the resources and with a high scientific return. Due to the large number of
parameters involved, the planning and scheduling of observations carried out by
human operators is a laborious and complicated process that does not guarantee
an optimal result.

The CARMENES GTO survey also includes a careful planning of the
observations to ensure the most efficient use of the telescope
time and thus to maximize the science output. Although most past surveys have
used a manual approach to planning, new projects increasingly appreciate the
importance of carefully optimizing the observation schedule. In this sense,
different mathematical tools to solve automated planning and scheduling
problems have been developed, ranging from simple heuristics to more complex
Artificial Intelligence applications \citep{Donati12, Policella13}.  Examples
include the scheduling tools of ALMA \citep{Espada14}, Las Cumbres Observatory Global Telescope \citep{Brown13},
\textit{EChO} \citep{Garcia-Piquer14d}, the Automated Planet Finder Telescope
\citep{Burt15}, MrSPOCK for the \textit{Mars Express} mission \citep{Cesta09}, the
\textit{James Webb Space Telescope} \citep{Giuliano11}, and the SOFIA mission
\citep{Civeit13}. To complement these examples, a summary of other planning
and scheduling tools used in astronomical observatories can be found in
\citep{Colome12}.

In particular, Genetic Algorithms (hereafter GAs), which are Evolutionary Computation
(EC) techniques \citep{Holland75b}, are very useful for this purpose. EC is an
Artificial Intelligence subfield focused on emulating natural evolution by means of
combining potential solutions using selection, combination and mutation operators
\citep{Freitas02}. The goal of GAs is to efficiently explore a large amount of
potential solutions in order to find near-optimal solutions \citep{Goldberg89}
fulfilling all constraints and optimizing the goals defined in the problem. GAs
must be adapted to the particularities of the problem in order
to obtain suitable results \citep{GarciaPiquer13}. Generally, a scheduler for
astronomical observations has more than one parameter that needs to be optimized,
resulting in a Multi-objective Optimization Problem (MOP) that can be defined as the
problem of finding a vector of decision variables satisfying constraints and
optimizing a vector function whose elements represent the objective functions
\citep{Osyczka1985}. These functions form a mathematical description
of performance criteria that are usually not disjoint (i.e., they are in conflict
with each other). Hence, the term ``optimize'' refers to finding a solution that
yields acceptable values for all objective functions \citep{Coello1999}.
Usually there is not a single point that simultaneously optimizes all
the objective functions of a MOP. Therefore, in these problems it
is necessary to look for trade-offs, rather than single solutions. The concept of
Pareto Optimality \citep{Pareto1897} defines that we can consider a Pareto optimal
when no feasible vector of decision variables exists that would decrease some
criterion without causing a simultaneous increase in at least one other criterion.
Thus, this concept almost always does not yield a single solutions but
a set of solutions called the Pareto optimal set. All solutions included in
the Pareto optimal set are non-dominated (i.e., there is no solution better
than the rest) and they have a different trade-off between objectives
\citep{GarciaPiquerPhd}.
The plot of the objective functions whose non-dominated
vectors are in the Pareto optimal set is called the Pareto front
\citep[see,][for further details]{Coello1999,Coello2001}.
Multi-Objective Evolutionary Algorithms \citep[hereafter, MOEAs][]{Coello2007}
are recognized as one of the most valuable and promising approaches to addressing
complex and diverse problems of multi-objective optimization.

In this paper we present in detail the scheduling tool developed in the context
of the CARMENES GTO survey, already introduced in \citep{Garcia-Piquer14}.
Although it has been designed for a radial-velocity application, it is seamlessly
adaptable to other purposes and constraints \citep[see
e.g.,][]{Garcia-Piquer14d}. In Sect.~\ref{sec:CAST} we describe the general
implementation of the scheduling tool, based on GAs. In
Sect.~\ref{sec:efficiency} we apply
the constraints of the CARMENES survey and we analyze the efficiency of the
scheduling algorithms in terms of telescope and instrument operations. Finally,
in Sect.~\ref{sec:science} we analyze the results of the optimal schedule
calculations to assess their scientific efficiency in the case of a radial
velocity survey and to estimate the CARMENES expected exoplanet yield. The
conclusions are presented in Sect.~\ref{sec:conclusions}.

\section{CAST}
\label{sec:CAST}

\begin{figure}[t]
    \center
   \includegraphics[width=0.5\textwidth]{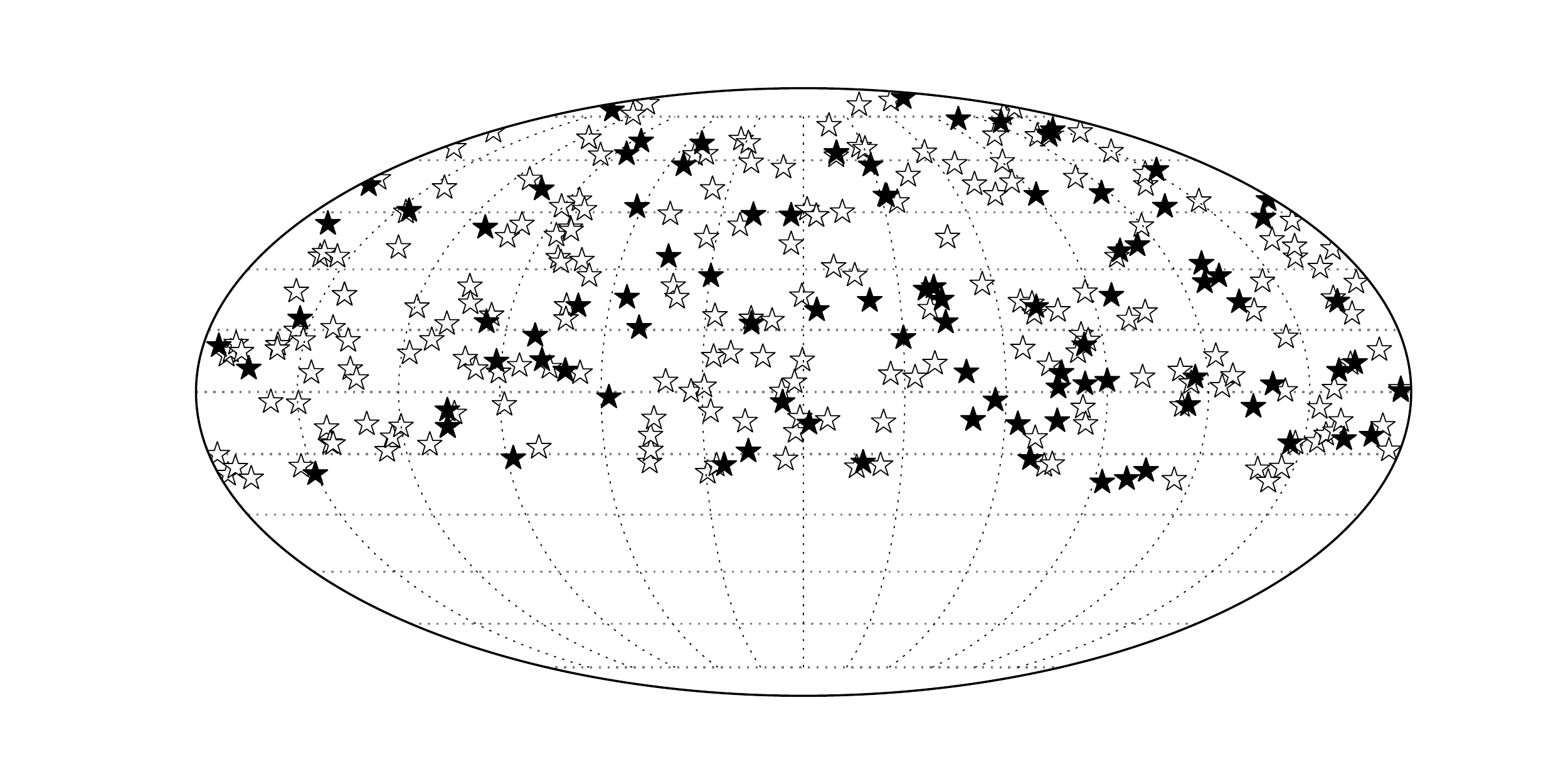}
   %\vspace*{-1.0cm}
   \includegraphics[trim = 0mm 15mm 0mm 32mm, clip=true, width=0.5\textwidth]{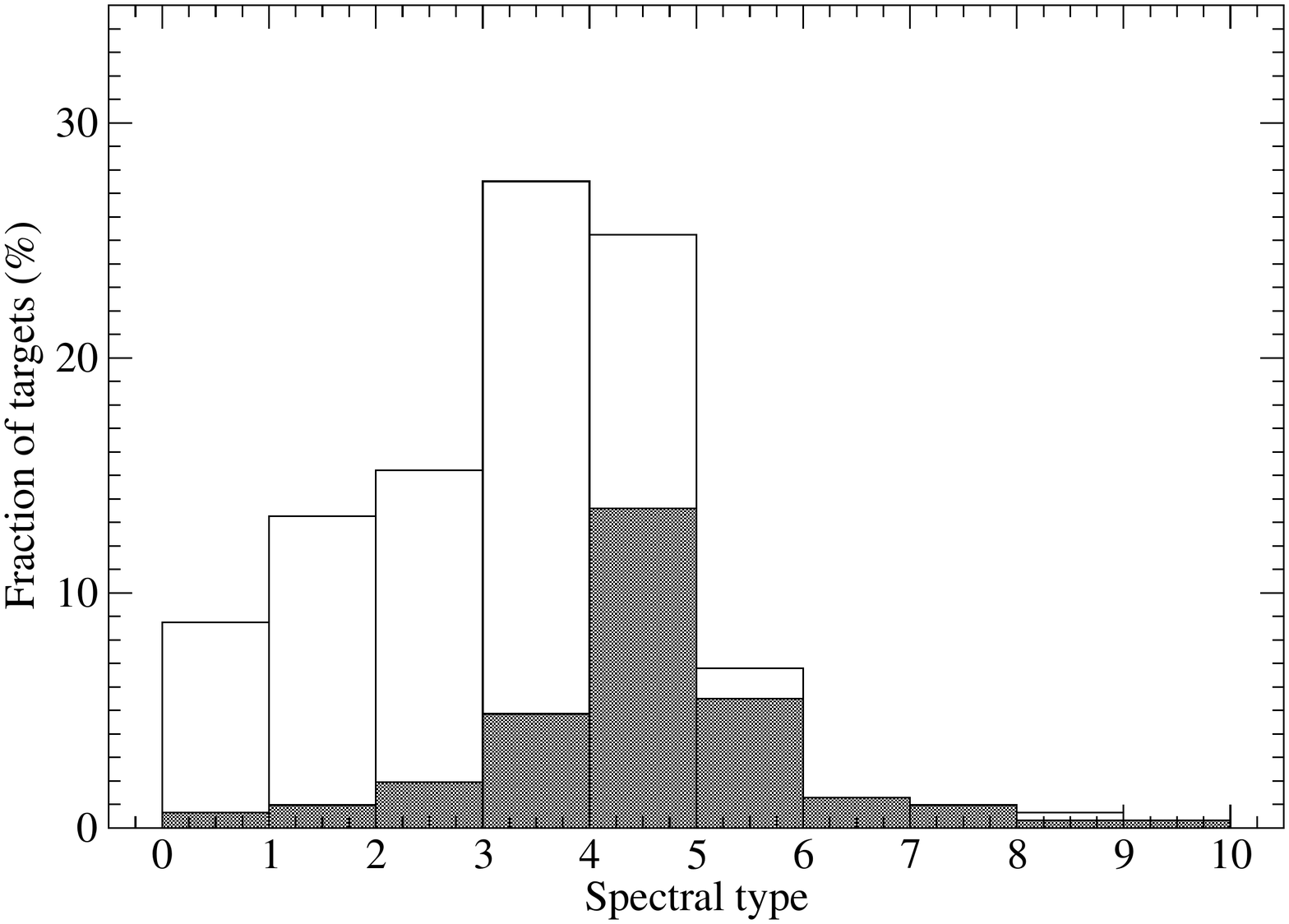}
   %\vspace*{-0.9cm}
    \caption{\textit{Top}: Mollweide projection of the distribution on the sky of the
309 CARMENES targets used in our simulations. \textit{Bottom}: Spectral type
distribution of the targets. In both panels, filled symbols and bars indicate
active stars with the H$\alpha$ line in emission.}
    \label{fig:targets}
\end{figure}

The first step to schedule an astronomical survey is the preparation of
the sample of targets and the identification of
observational constraints that may be defined by the target stars, the
telescope and instrument design, and the science requirements.
As mentioned above, CARMENES will carry out a
survey of $\sim$300 M dwarfs for a period of at least three years. For this paper,
we have used a list of 309 high-priority potential target candidates compiled in
the CARMENES input catalogue \citep[dubbed Carmencita;][]{Caballero13a,AlonsoFloriano15}.
The CARMENES science objectives  for at least 60 observations to be obtained for
each target. These high-priority M dwarfs are selected from a larger sample
by removing unsuitable systems for the survey (e.g., faint stars in the
$J$-band for their spectral subtypes, or spectroscopic or close visual binaries). The list comprises
targets distributed among all M dwarf spectral subtypes as shown in
Fig.~\ref{fig:targets}, which also depicts the distribution of targets on the
sky. The $J$ magnitude ranges from 4.2 to 11.2~mag, with a mean value of
7.7$\pm$1.0~mag. For simplicity, the simulation experiments presented in
this paper assume that all targets have the same priority. However, an
external priority value is included in the description of the algorithm.
This is because such functionality has been implemented in the operational
CARMENES scheduler in case that it needs to be considered for the optimization
over the course of the survey (e.g., to increase measurement cadence of
interesting targets or to sample specific periods). This priority is an
integer value that is assigned to each target by a user, with larger
values indicating higher priority.

Regarding the constraints of the survey, they can be
divided into two different categories: hard constraints and soft constraints.
The first ones have to be necessarily satisfied, and the second ones express a
preference of some target combinations over others. Thus, the final scheduling
solution must fulfill all hard constraints and should optimize soft
constraints.

\subsection{Scheduling constraints}
\label{subsec:CASTconstraints}

In the case of the CARMENES GTO survey, the hard constraints implemented in the
CARMENES Scheduling Tool (hereafter CAST) are mainly related to the visibility
of the targets from the Calar Alto Observatory, the operation overhead times,
and the environmental conditions. These restrictions can also be adapted for
other surveys and facilities such as space-based telescopes \citep[see
e.g.,][]{Garcia-Piquer14d}. The \emph{hard constraints} identified are
summarized in Table~\ref{tab:constraints} and are described as follows:

\begin{table}[t]
\caption{List of CARMENES observational constraints.}
\vspace*{-0.35cm}
\begin{center}
\label{tab:constraints}
\begin{footnotesize}
\begin{tabular}{lll}%p{1.2cm}}
\hline
\hline
\noalign{\smallskip} 
\multicolumn{1}{l}{{Constraint}} & \multicolumn{1}{l}{{Category}} & \multicolumn{1}{l}{{Computation}} \\
\noalign{\smallskip} 
\hline
\noalign{\smallskip} 
 Night 	&	Hard 	&	 In advance 	\\ %\hline
 Elevation	&	Hard 	&	 In advance 	\\ %\hline
 Moon influence 	&	Hard 	&	  In advance 	\\ %\hline
 Visibility duration 	&	Hard 	& In advance 	\\ %\hline
 Pointing 	&	Hard 	&	  In advance 	\\ %\hline
 Overlapping 	&	Hard 	&	  In advance \\ %\hline
 Overhead time 	&	Hard 	&	  In advance \\ %\hline
 Environmental conditions 	&	Hard 	& On the fly \\ %\hline
 Observing time 	&	Soft 	& In advance \\
                        &               & On the fly \\ %\hline
 Observation deviation 	&	Soft 	& In advance \\ %\hline
 Observing cadence 	&	Soft 	& In advance \\ 
\noalign{\smallskip} 
 \hline
\end{tabular}
\end{footnotesize}
\end{center}
\end{table}

\begin{enumerate}
\item \emph{Night. }
The object shall only be observed from afternoon twilight to morning twilight.
The coordinates of the targets on the sky and twilight times are computed
according to the date of observation and the location of the observatory.
Additionally, if needed, the observability window of each target can be
computed according to given ephemerides for the case of objects with periodic
variability. { For this paper we have considered the start of the astronomical
twilight at Calar Alto Observatory as a conservative approach (Sun at $-12$\,deg
elevation).}
\item \emph{Elevation. }
The elevation of each object is calculated according to its equatorial coordinates
and the geographic
coordinates of the observatory. The objects shall only be observed if they
exceed a specific elevation for at least a certain amount of time. The
elevation and time are two parameters that can be introduced in the global
configuration of CAST. { For CARMENES, the scheduling requirement is
that a target must be above 30\,deg for a time span longer than its estimated
integration time .}
\item \emph{Moon influence. }
Targets shall be observed when \emph{1)} the Moon is below the horizon or
\emph{2)} the Moon is sufficiently far so that the observation is not
significantly contaminated by background light. { We have established that the
minimum acceptable distance to the Moon ($r_{\rm min}$) is 20\,deg. Beyond
this distance, a hard constraint function on the Moon is evaluated to select
only targets that are at least 5 magnitudes brighter than the background (see
Appendix~\ref{ap:moon} for further details).}
\item \emph{Visibility duration. }
The total time during which the \emph{Night}, \emph{Elevation} and \emph{Moon
influence} constraints are fulfilled shall be equal or higher than the minimum
visibility time required for a target observation. { This minimum time
corresponds to the exposure time, which is computed using a calibration as a
function of $J$-band magnitude using real CARMENES observations
(see Appendix~\ref{ap:integrationtime} for more details). 
The maximum exposure time is set to 30 minutes to avoid
biasing the barycentric correction.
This exposure time limitation only affects 35 late-type dwarfs in our sample,
which would need longer integration times according to their $J$-band magnitude.
However, the effect of the reduction of exposure time is counterbalanced by the
fact that for faint targets, typically late M-dwarf stars, the radial-velocity
precision needed to detect planets is not as high as for earlier types.}
\item \emph{Pointing. }
In case of pointing restrictions, targets shall only be observed if they are
between minimum and maximum elevations as defined by the survey requirements.
Moreover, in the case of CARMENES, the telescope dome has a ``segmented'' hatch
that allows five open window configurations with different apertures (see
Appendix~\ref{ap:point} for more details). If the window
configuration needs to be changed (because of vignetting) during target
integration, the slide would temporarily block the telescope aperture and the
observation will be affected. Thus, an observation shall only be selected if
the target can be observed without being obstructed for the entire duration of
the estimated integration time.
\item \emph{Overlapping. }
In operational terms, there are three kinds of tasks to be considered:
\emph{1)} target observation, \emph{2)} read-out of the previous observation, and
\emph{3)} slewing to acquiring a new target. Only tasks \emph{2)} and \emph{3)}
can be executed in parallel.
\item \emph{Overhead time. }
Pointing to a particular object requires a specific telescope and instrument
configuration. The time between consecutive observations considers both the
telescope slew \& acquisition time and the instrument read-out time. The
former includes the time needed to move the dome, the hatch and the telescope,
and an overhead slew time for acquisition, while the latter is defined by the
detector properties. We assume the total overhead time between observations
to be the duration of the process that takes a longer time (see
Appendix~\ref{ap:overheadTime} for the detailed logical operations). { For the
CARMENES scheduler, we assume a telescope and dome slew rate of 1\,deg per second,
and 60\,s to change the dome hatch (the segments can be moved together, so the time
overhead is independent of the number of segments to move). The overhead slew time
is 120\,s and the detectors read out time is less than 40\,s.}
\item \emph{Environmental conditions. }
An observation can be programmed when the environmental conditions permit. In
operation mode during the survey, the CARMENES Instrument Control System
\citep{Garcia-Piquer14} informs CAST if the environmental conditions are
suitable for observation. In the simulation mode presented here, we have used a
weather model based on Calar Alto environmental conditions from 2004 to 2006
(Calar Alto Observatory, priv. comm.), {thus taking into account seasonal
weather influence}. We assumed that the dome is closed
if: \emph{1)} relative humidity reaches 98\,\%, and must be closed until it is
equal to or below 95\,\% for at least 20 minutes; \emph{2)}
the outside temperature is below $-$15\,$\degree$C; or \emph{3)} the wind speed is
above 24~\ms\/. Furthermore,
{to be conservative, we assume that a given night has a maximum probability
of 20\,\% of cloudiness in low humidity conditions or of technical issues impeding
observations.}
And, finally, the simulator increases the integration time randomly up to
20\,\% to simulate the effect of high clouds.
\end{enumerate}

On the other hand, in terms of scheduling, science requirements are identified
as \emph{soft constraints}. These restrictions are also summarized in
Table~\ref{tab:constraints} and include:
\begin{enumerate}
\item \emph{Observing time. }
The integrated global observing time (i.e., the time that the telescope is
observing), especially that of high-priority objects, should be maximized. This
guarantees that the most interesting targets are sufficiently observed.
\item \emph{Observation deviation. }
The variance of the number of times that objects of the same priority have been
observed in the complete survey should be minimized. This constraint should
ensure that all targets will have a proper share of assigned observing time.
\item \emph{Observing cadence. }
It is possible to select the number of observations per night required for each
target. As an optional constraint, the planning tool includes functionality to
observe the targets at appropriate times to increase planet detectability by,
e.g., averaging the intrinsic stellar noise on radial velocities \citep[see,
e.g.,][]{Dumusque11b,Dumusque11a} or by optimizing the periodogram window
function over a certain orbital period interval (i.e., avoiding peaks or gaps).
In case of transient objects, target ephemerides are considered.  For the
present work, we assume that each target will be observed no more than once
per night and randomly for the duration of the survey.
\end{enumerate}

Some of the constraints can be computed in advance but others, such as weather
conditions (environmental conditions), can only be determined in real time
during observations and the scheduler must be reactive to their variations.
For this reason, although being an independent system, CAST is connected with
the Instrument Control System from which it receives environment parameters
and return an observation request optimized according to current conditions. In
order to reduce waiting times, one of the CAST requirements is that it should
invest less than five seconds in selecting the next target to be observed. The
simultaneous fulfilment of the hard constraints and the optimization of the
soft constraints should provide a scheduling solution that maximizes the
scientific return of the survey.

\subsection{Scheduling optimization} \label{subsec:CASToptimization}

Given the large number of targets of the CARMENES sample and its conditions,
the complexity in computing the enormous amount of possible combinations in
search for a near-optimal scheduling solution is unaffordable for human
operators. Hence, the scheduling of CARMENES observations can be addressed as a
constraint-satisfaction problem, which is a combinatorial problem that seeks an
assignment of values to its variables that will satisfy all given constraints.
The algorithms used to find the optimal solution to these kind of problems
require a computation time that grows exponentially with the size of the input.
Consequently, approximate or heuristic methods are useful to find feasible
solutions, or those that satisfy most of the constraints, in a reasonable
computation time \citep{Nonobe98}. { In CAST, GA techniques are implemented
in order to efficiently explore the large amount of potential combinations
of observations with the goal of selecting those that are more efficient.}

\begin{figure*}[!t]
    \center
   \includegraphics[trim = 2mm 66mm 5mm 0mm, clip=true, width=0.6\textwidth]{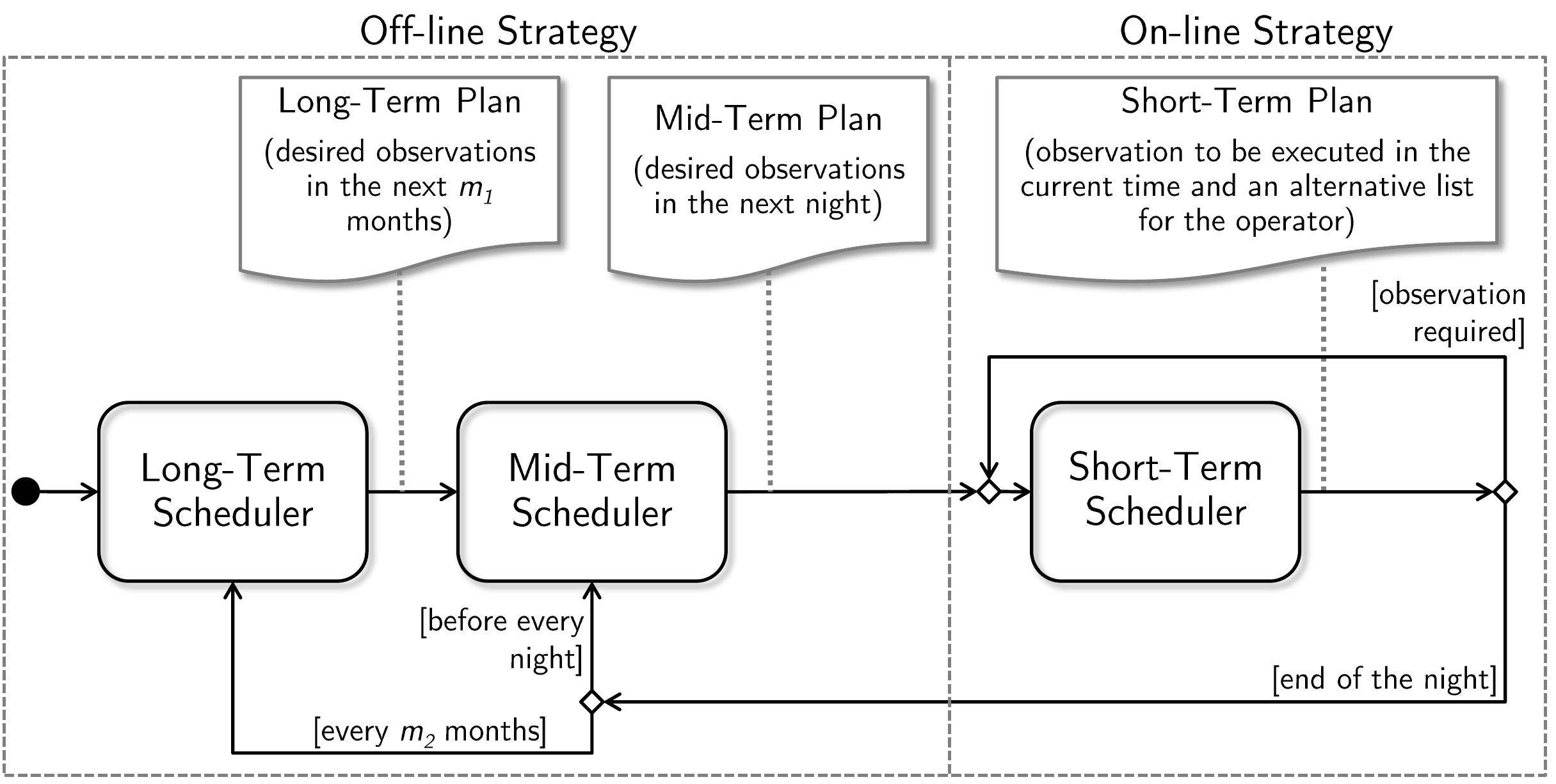}
    %\vspace*{0.2cm}
    \caption{Combination of the three schedulers of CAST. The Long-term
scheduler plans observations with a time scope of several months ($m_1$) and is
recomputed after a specific number of months ($m_2$). The Mid-term scheduler is
executed daily and computes the observations to be performed during each night.
The Short-term scheduler is executed during the night after every exposure
and computes the next optimal observation and an alternative ranked list of alternatives. In this
work, we set $m_1$=6 and $m_2$=3\,months.}
    \label{fig:schedAct}
\end{figure*}

%{ CAST is based on Genetic Algorithms (GAs), which are
%Evolutionary Computation (EC) techniques \citep{Holland75b}.}
The algorithms of CAST were presented in \citep{Garcia-Piquer14} but we provide
the basic details here. Two scheduling strategies are included in CAST:
off-line and on-line \citep{Rasconi06a}. The off-line strategy includes two
planning tools, Long-term and Mid-term schedulers, which are designed to plan
the targets to be observed within a time interval according to the hard
constraints that can be predicted. The on-line strategy considers a Short-term
scheduler that takes into account all constraints and adapts the previously
computed mid-term plan to the immediate circumstances \citep{Rasconi06b}. Table
\ref{tab:constraintsScope} summarizes the constraints considered by each of the
three schedulers. Their combination is illustrated in Fig.~\ref{fig:schedAct}
and we provide additional details on the three scheduling tools in
Sect.~\ref{subsec:CASTGAOff}.
%\ref{s:lt_sched}, \ref{s:mt_sched}, and \ref{s:st_sched}.

\begin{table}[!t]
\caption{Constraints used by the CAST Long-, Mid-, and Short-term schedulers.}
\vspace*{-0.35cm}
\begin{center}
\label{tab:constraintsScope}
\begin{footnotesize}
\setlength\tabcolsep{2.3pt}
\begin{tabular}{lccc}
\hline
\hline
\noalign{\smallskip} 
Constraint	&	Long-term	&	Mid-term	&	Short-term	\\
\noalign{\smallskip} 
\hline
\noalign{\smallskip} 
Hard constraints & & & \\
\noalign{\smallskip} 
\hline
\noalign{\smallskip} 
 Night 	&	$\surd$	&	$\surd$	&	$\surd$	\\
 Elevation	&	$\surd$	&	$\surd$	&	$\surd$	\\
 Moon influence 	&	$\surd$	&	$\surd$	&	$\surd$	\\
 Visibility duration 	&	$\surd$	&	$\surd$	&	$\surd$	\\
 Pointing 	&	$\surd$	&	$\surd$	&	$\surd$	\\
 Overlapping 	&		&	$\surd$	&	$\surd$	\\
 Overhead time 	&		&	$\surd$	&	$\surd$	\\
 Environmental conditions 	&		&		&	$\surd$	\\
% System Conditions 	&		&		&	$\surd$	\\
\noalign{\smallskip} 
\hline
\noalign{\smallskip} 
Soft constraints & & & \\
\noalign{\smallskip} 
\hline
\noalign{\smallskip} 
 Observing time 	&		&	$\surd$	&	$\surd$	\\
 Observation deviation 	&		&	$\surd$	&	$\surd$	\\
 Observing cadence 	&	$\surd$	&		&		\\ \hline
\end{tabular}
\end{footnotesize}
\end{center}
\end{table}

\subsection{Genetic Algorithm for the off-line strategy}
\label{subsec:CASTGAOff}
The Long-term and the Mid-term schedulers use a MOEA to plan the science
observations by promoting the soft constraints defined for each scheduler.
The MOEA process, as a GA, is generally based on selection, reproduction, and
mutation processes. This paradigm makes it possible to explore all regions of
the parameter space, which is a vast area with a large amount of potential
solutions, in search of the best solutions. This kind of algorithms begins with
a set of initial solutions that are improved through an iterative cycle based
on evaluating, selecting, recombining, and mutating them. The key aspect for
finding high quality solutions lies in the definition and design of individual
representation, the genetic operators, and the objectives to be optimized
\citep{Goldberg02, GarciaPiquerPhd}. For more details about the design of a
MOEA, the reader is referred to Appendix~\ref{ap:moea}.
The MOEA applied is based on the NSGA-II procedure \citep{Deb2002}, which is
one of the most well-known MOEAs.
However, with the aim of obtaining a suitable optimization solution, it is
necessary to design ad hoc parts of the algorithm according to the problems
to be solved, so the Long-term and Mid-term schedulers have some differences.

\subsubsection{Long-term scheduler}\label{s:lt_sched}
The Long-term scheduler plans object observations with a time scope of several
months. It takes into account the \emph{Night}, \emph{Elevation}, \emph{Moon
influence} and \emph{Visibility duration} constraints. A procedure based on MOEAs
is applied to identify the best nights when each object should be observed by
optimizing the soft constraint regarding the \emph{Observing cadence}
constraint. The resulting plan provides a list of potential observation dates
for each object assuming that all scheduled nights for the survey are suitable.
Besides, the Long-term scheduler is run periodically to take into account
observations previously done, thus counteracting any effect that unpredicted
bad weather may have on the optimization. The execution of this scheduler is
not time-critical because it is run daily before the start of telescope
operations and without interaction with the Instrument Control System to
retrieve weather information. Therefore, it can be used as an standalone
planning tool for any observatory.

The design of the MOEA used in the Long-term scheduler is defined as follows:
\begin{itemize}
\item
%Individual representation.
%In the Long-term scheduler, the proposed
The individual genotype uses a binary
encoding that represents if the target is planned on a given night. Each
individual consists of $\overline{\overline{N}}$ genes $\{o_{1}, \cdots ,
o_{\overline{\overline{N}}}\}$, where $\overline{\overline{N}}$ is the
cardinality of the set of
nights ($N$) that the Long-term scheduler considers (e.g., the number of nights
in three months) and $o_{i}$ corresponds to night $i$. Moreover, the $o_{i}$
value has to be 0 or 1, where 0 indicates that the target is not planned in the
corresponding night and 1 indicates that the target is planned. The order of
the targets in the genotype indicates a temporal sequence from the first night
considered to the last one. The initial population is built by creating $N_I$
new individuals assigning to each allele a value of 0 or 1 with a probability
of 0.5. This representation does not allow unfeasible individuals.

\item
%Genetic operators. \label{sec:gaLT}
The selection, crossover and mutation operators are described in Appendix~\ref{ap:genops}. 
In the case of the Long-term scheduler, a mutated gene $g'$ is
obtained by negating $g$ (i.e., alleles that are 0 become 1, and those that are
1 become 0).

\item
%Objective functions.
The Long-term scheduler has the optimization goal of
identifying the nights when an object should be observed according to the
\emph{Observing cadence} constraint. In the case of CARMENES, it is desirable
to observe the targets when they are near their culmination. 
The optimization objectives promote the observation of an object near
meridian crossing at the same time that maximizes the number of observations done.
Thus, two objective functions are defined: $F_{\rm c}$ promotes the observations
of an object near meridian crossing and $F_{\rm n}$ promotes the number of
observations of the object, as described by Eqs.~\ref{eq:longFitness1} and
\ref{eq:longFitness2}, respectively.
Where {$O$ is the set of nights when target $t$ is observed, $N$ is the set of
genes of the individual chromosome (i.e., the number of nights to be planned)},
$Z_{\rm min}(t)$ computes the minimum zenith angle that target $t$ can achieve,
$Z_{\rm min}(t, n)$ computes the minimum zenith angle that target $t$ can
achieve in night $n$,
and $\overline{\overline{O}}$ and $\overline{\overline{N}}$ are the numbers of
elements (cardinality) in sets $O$ and $N$, respectively
(i.e., the number of nights when a target is observed and the total number of
nights in the planning). 
$F_{\rm c}$ and $F_{\rm n}$ have normalized values between 0 and 1,
and they are optimized when minimized.

\begin{tiny}
\begin{align}
\label{eq:longFitness1}
\begin{split}
%&F_{\rm c}\left(N, t\right){\rm =}1-\left(\frac{\sum_{\rm n\in N}{culm(n, t)}}{\overline{\overline{N}}}\right)\ ,\\
&F_{\rm c}\left(O, t\right){\rm =}1-\frac{1}{\overline{\overline{O}}}\cdot\left(\sum_{\rm o\in O}{\frac{Z_{\rm min}(t)}{Z_{\rm min}(t, o)}}\right)\ . 
%&culm\left(n, t\right){\rm =}\frac{Z_{\rm min}(t)}{Z_{\rm min}(t, n)}\ .
%&culm\left(n, t\right){\rm =}\frac{ams(globalmaxalt(t))}{ams(maxalt(n, t))}\ ,\\
%&ams\left(a\right){\rm =}\frac{1}{\cos((90 - a)\cdot(\frac{\pi}{180}))}\ .
\end{split}
\end{align}
\end{tiny}
\begin{tiny}
\begin{align}
\label{eq:longFitness2}
\begin{split}
&F_{\rm n}\left(N, O\right){\rm =}1-\left(\frac{\overline{\overline{O}}}{\overline{\overline{N}}} \right)\ .
\end{split}
\end{align}
\end{tiny}

\item
%{Selection of the most suitable solution}
We consider that the most suitable solution is the individual that has the
lower average of objectives (the defined objectives have to be
minimized to be optimized) from all the individuals in the first Pareto
front (i.e., non-dominated individuals).

\end{itemize}

\subsubsection{Mid-term scheduler}\label{s:mt_sched}

The Mid-term scheduler plans the observations that should be executed during a
specific night by optimizing the \emph{Observing time} and \emph{Observation
deviation} soft constraints, and according to the results of the long-term plan.
Moreover, the resulting mid-term plan fulfills all the hard constraints that are
predictable: \emph{Night}, \emph{Elevation}, \emph{Moon influence},
\emph{Visibility duration}, \emph{Pointing}, \emph{Overlapping}, and
\emph{Overhead time}. The execution of this scheduler is not
time critical because it can be run before the start of the nightly telescope
operation, and therefore, a GA is used to obtain a near-optimal plan.

The MOEA used in the Mid-term scheduler is designed as follows:
\begin{itemize}
\item
%Individual representation
The proposed individual genotype is made up of
double numbers that represent the starting time of the observation of the
targets. Each individual consists of $\overline{\overline{T}}$ genes $\{o_{1},
\cdots, o_{\overline{\overline{T}}}\}$, where $\overline{\overline{T}}$ is
the cardinality of the set of
targets to be planned ($T$), and $o_{i}$ corresponds to target $i$. Moreover,
the $o_{i}$ value has to be between the range [${w}_{st}$, ${w}_{et} - d_t$],
where $w$ is a random uniform window in $W_t$, which are all the visibility
windows in the night for target $t$, ${w}_{st}$ is the Julian day of the
starting time of window $w$ for target $t$, ${w}_{et}$ is the Julian day of
the ending time of window $w$ for target $t$, and $d_t$ is
the estimated integration time in Julian days for target $t$ (see
Appendix~\ref{ap:integrationtime}). Moreover, $o_{i}$ can have a value of $-1$
indicating that target $i$ does not have a starting time
assigned (i.e., it is not planned). The order of the targets in the genotype
does not indicate a temporal sequence, but it is only the order of the targets in
the input data. The temporal sequence of targets is defined by the alleles,
because they indicate the starting time assigned to each target. For instance,
a target in position $i$ of the genotype can be planned in a time window
previous to the time window of target $i-1$. The initial population is built
by creating $N_I$ new individuals assigning to each allele $o_{i}$ a $-1$ value
or a value in the range between [${w}_{st}$, ${w}_{et} - d_t$] following a uniform
distribution. The process to build each individual is based on placing the
observations of the targets, selected in random order, and avoiding overlaps.
In case of overlapping, the target is unplanned (i.e., a value $-1$ is assigned).
\item
%Genetic Operators
Selection, crossover and mutation operators work as Appendix~\ref{ap:genops} shows. 
In the Mid-term scheduler, the mutation operator alters a gene $g'$ by changing
its allele with a value inside the potential time
windows of the corresponding target. Thus, a mutated gene $g'$ changes its
allele with a random uniform value $\mu$ in the range [${w}_{st}$,
${w}_{et} - d_t$] and $-1$  (i.e., $g' = \mu$).
In this case, the crossover of two feasible
individuals can generate unfeasible offspring due to overlapping, and the
mutation of a feasible individual can also generate an unfeasible solution.
This is solved by a repairing procedure devoted to obtain feasible new
individuals, as the next point explains.
\item
%Repair of the Individual
An individual represents the time windows assigned to target observations, but
it does not consider the slew time between two observations. Thus, this aspect
has to be considered for obtaining the final planning codified by each
individual. This modification can produce an unfeasible individual because it
can have conflicting observations (i.e., presence of overlaps in the
observations). There are two ways for obtaining an unfeasible individual that
requires repair during the GA process: {\emph1)} the individual has overlapping
between two or more observations, and {\emph2)} there is overlapping between
two or more observations when slew time is added to each observation. We may
find that it is necessary to repair the individuals after the mutation process
in order to obtain feasible ones. Thus, the main idea of the repair operator is
to solve all overlaps in the individual by unplanning conflicting targets.
The unplanning of one target can solve overlaps between several
targets.
\item
%Objective functions
The optimization goal of the Mid-term scheduler is to plan the selected
objects according to two objectives related to the \emph{Observing time} and
\emph{Observation deviation} soft constraints, with the aim of minimizing the
instrument idle time (time of the night during which the instrument is not
acquiring scientific data) weighted with the priority completeness of the
targets, and mitigating the problem of scheduling the objects that
require longer observations. Consequently, the Mid-term scheduler optimizes
two functions. $F_{\rm w}$ promotes the time scheduled for observations of
objects near meridian crossing according also to their priority, and
$F_{\rm d}$ promotes a proper distribution of the observations of the objects
with the same priority, as described in Eqs.~\ref{eq:midFitness1} and
\ref{eq:midFitness2}, respectively.
Where $S$ is the set that contains the observations scheduled in the night,
$o_{\rm target}$ is a target associated to observation $o$, $o_{\rm datetime}$
is the mid time of observation $o$, $integrationTime(o_{\rm target})$ computes
the estimated integration time in seconds of a target observation,
$night duration$ indicates the duration of the night in seconds,
$priority(o_{\rm target})$ is the priority of the target associated to
observation $o$ normalized according to the priorities of all targets,
%$t_{\rm priority}$ is the priority of target $t$, with the largest values indicating the highest priority,
%$P$ is the set of priorities,
$Z_{\rm min}(o_{\rm target})$ computes the minimum zenith angle that target $o_{\rm target}$ can achieve
during the night,
$Z_{\rm min}(o_{\rm target},o_{\rm datetime})$ computes the minimum zenith angle of target $o_{\rm target}$ at the date
and time $o_{\rm datetime}$,
{$P$ is the set of priorities},
$S_{\rm p}$ is the set that contains the observations of targets with priority
$p$ in the night, $T_{\rm p}$ is the set of the targets with priority $p$,
$S_{\rm t}$ is the set that contains the observations of target $t$,
and $\overline{\overline{P}}$, $\overline{\overline{S}}_{\rm p}$,
$\overline{\overline{T}}_{\rm p}$, $\overline{\overline{S}}_{\rm t}$ are the
number of elements (cardinality) in sets $P$, $S_{\rm p}$, $T_{\rm p}$, and
$S_{\rm t}$, respectively. $F_{\rm w}$ and $F_{\rm d}$ have values
between 0 and 1, and they are optimized when minimized.

\begin{tiny}
\begin{align}
\label{eq:midFitness1}
\begin{split}
&F_{\rm w}\left(S\right){\rm =}1-\left(\frac{\sum_{\rm o\in S}{\left(integrationTime(o_{\rm target})\cdot weight(o)\right)}}{night\ duration}\right)\ ,\\
&weight(o){\rm =}priority(o_{\rm target}) \cdot \frac{Z_{\rm min}(o_{\rm target})}{Z_{\rm min}(o_{\rm target},o_{\rm datetime})}\ .
%&weight(o){\rm =}priority(o_{\rm target}) \cdot culm(o_{\rm target}, o_{\rm datetime})\ ,\\
%&priority(t)=\frac{t_{\rm priority}+1}{\overline{\overline{P}}+1}\ ,\\
%&culm(t,dt){\rm =}\frac{Z_{\rm min}(t)}{Z_{\rm min}(t,dt)}\ .
\end{split}
\end{align}
\end{tiny}
\begin{tiny}
\begin{align}
\label{eq:midFitness2}
\begin{split}
%&F_{\rm d}\left(S\right)=\frac{\sum_{\rm p\in P}{stdev\left(S,p,\overline{p}\right)}}{\overline{\overline{P}}}\ ,\ \ \ \ {\rm where}\ \overline{p}{\rm =}\frac{\overline{\overline{S}}_{\rm p}}{\overline{\overline{T}}_{\rm p}}\ ,\\
&F_{\rm d}\left(S\right){\rm =}\frac{\sum_{\rm p\in P}{stdev\left(S,p,\overline{p}\right)}}{\overline{\overline{P}}}\ ,\\
&\overline{p}{\rm =}\frac{\overline{\overline{S}}_{\rm p}}{\overline{\overline{T}}_{\rm p}}\ ,\\
&stdev\left(S_{\rm p},p,\overline{p}\right){\rm =}\sqrt{\sum_{\rm t\in T_{\rm p}}{\frac{{(\overline{\overline{S}}_{\rm t}-\overline{p})}^2}{\overline{\overline{T}}_{\rm p}-1}}}\ .
\end{split}
\end{align}
\end{tiny}

\item
%Selection of the most suitable solution
The same strategy as in the Long-term scheduler is applied to select the most
suitable solution (see Sect.~\ref{s:lt_sched}).
\end{itemize}

\subsubsection{Short-term scheduler}\label{s:st_sched}

The Short-term scheduler computes the next observation to be executed during
the night by optimizing the \emph{Observing time} and \emph{Observation
deviation} soft constraints and by considering all previous observations. Moreover,
the selected observation fulfills all the hard constraints, Thus, this scheduler
reacts to immediate conditions (weather, errors, delays, events). Unlike
the Long-term and Mid-term schedulers, the Short-term scheduler is time
critical because it has to select an observation in a short time.
%(with a requirement less than 5\,s in the case of CARMENES). 
For this reason, in order to avoid intensive calculations, it repairs
the night schedule obtained by the Mid-term scheduler \citep{Akturk99}
using astronomy-based heuristics \citep{Giuliano07} instead of using a GA.
The Short-term algorithm is called after the end of an observation, and its
process is explained below.

\begin{figure}[!b]
\begin{center}
\begin{tiny}
\centering
\begin{tabular}{p{1cm} l}
%\begin{subfigure}{1.0\columnwidth}
 \centering
 \vspace*{-2.3cm}(a) & \includegraphics[width=0.79\columnwidth]{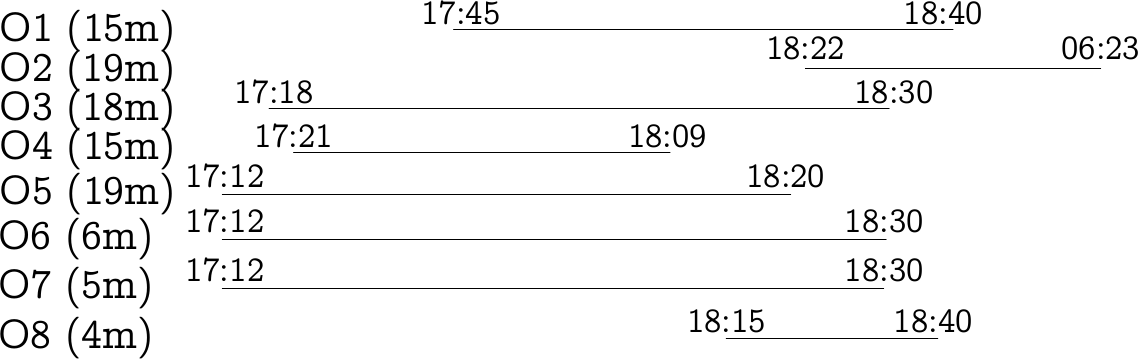}%~~~~~~~~~%this space is important to line up the temporal line
 \\
 %\caption{}%{Target visibility}
 \label{fig:st_1}
%\end{subfigure}
%\begin{subfigure}{1.0\columnwidth}
 \centering
 \vspace*{-1.2cm}(b) & ~~~~~~~~~~\includegraphics[trim = 210mm 200mm 95mm 75mm, clip=true, width=0.7\columnwidth]{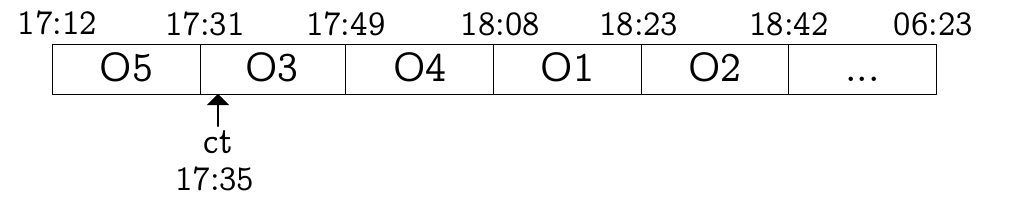}
 \\
 %\caption{}%{Observation selection from mid-term plan}
 \label{fig:st_2}
%\end{subfigure}
%\begin{subfigure}{1.0\columnwidth}
 \centering
 \vspace*{-1.2cm}(c) & ~~~~~~~~~~\includegraphics[trim = 210mm 170mm 95mm 105mm, clip=true, width=0.7\columnwidth]{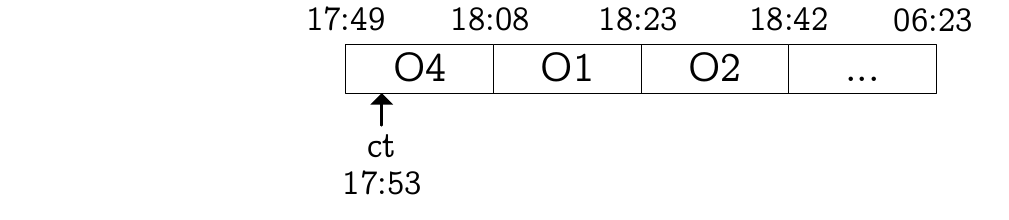}
 \\
 %\caption{}%{Observation selection from mid-term plan}
 \label{fig:st_3}
%\end{subfigure}
%\begin{subfigure}{1.0\columnwidth}
 \centering
 \vspace*{-1.2cm}(d) & ~~~~~~~~~~\includegraphics[trim = 210mm 135mm 95mm 140mm, clip=true, width=0.7\columnwidth]{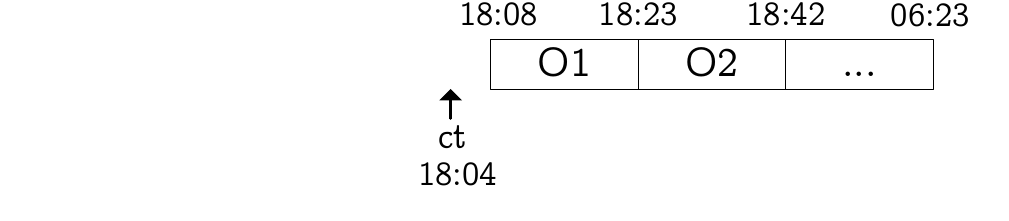}
 \\
 %\caption{}%{Observation selection from mid-term plan}
 \label{fig:st_4}
%\end{subfigure}
%\begin{subfigure}{1.0\columnwidth}
 \centering
 \vspace*{-1.2cm}(e) & ~~~~~~~~~~\includegraphics[trim = 210mm 107mm 95mm 168mm, clip=true, width=0.7\columnwidth]{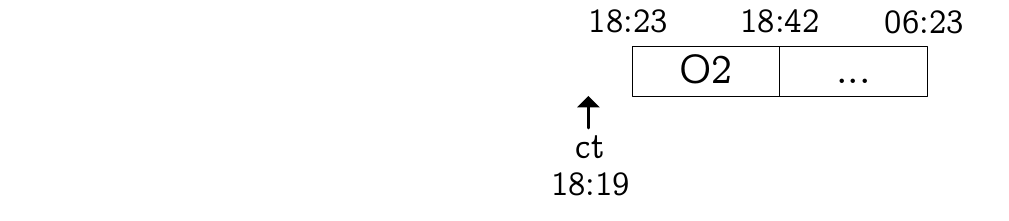}
 \\
 %\caption{}%{Observation selection from mid-term plan}
 \label{fig:st_5}
%\end{subfigure}
%\begin{subfigure}{1.0\columnwidth}
 \centering
 \vspace*{-1.2cm}(f) & ~~~~~~~~~~\includegraphics[trim = 210mm 81mm 95mm 194mm, clip=true, width=0.7\columnwidth]{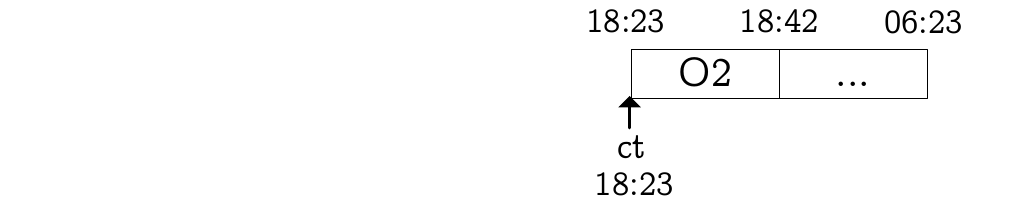}
 \\
 %\caption{}%{Observation selection from mid-term plan}
 \label{fig:st_6}
%\end{subfigure}
%\begin{subfigure}{1.0\columnwidth}
 \centering
 \vspace*{-0.6cm}(g) & ~~\includegraphics[trim = 198mm 65mm 113mm 220mm, clip=true, width=0.66\columnwidth]{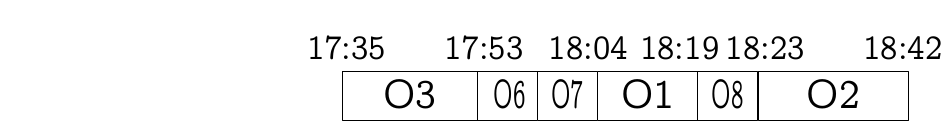}
 \\
 %\caption{}%{Executed schedule}
 \label{fig:st_7}
%\end{subfigure}
\end{tabular}
\end{tiny}
\caption{{
Examples of observation selection corresponding to time $ct$. $Ot$
indicates an observation of target $t$. For simplicity, the slew time between
observations is not considered. (a) represents the visibility (i.e., fulfillment
of the hard constraints) and integration time of each target in minutes. The line indicates
the time window where the target is visible, and the integration time of the
observations in minutes is indicated in parentheses. (b), (c), (d), (e), and (f)
present the objects planned in the mid-term plan that are available at $ct$. (b),
shows how the Short-term scheduler discards $O5$ because it is not active in the mid-term plan at
$ct$. Thereafter, $O3$ will be shifted and will start at 17:35, because the observation
will end inside its visibility period as it is visible until 18:30. Next, in (c), $O4$ is
not visible after 18:09 and it cannot be completed if it is delayed until 17:53. Thus,
$O4$ is removed from mid-term plan. Consequently, there is a gap between 17:53 and
18:08 that will be filled with one or several new observations. In this case, we assume
that two new observations are executed ($O6$, $O7$). (d) depicts how the gap is only
filled until 18:04 because no visible target that is not already in the mid-term plan
fits in the remaining window. $O1$ is advanced and it will start at 18:04 because it
is visible at this time. In (e), $O2$ is not visible until 18:22, so it cannot be
advanced. The gap between 18:19 and 18:23 will be filled with one or several new
observations. In this case, we assume that one new observation is executed ($O8$).
Finally, in (f), $O2$ can start at the time scheduled by the mid-term plan.
(g) shows the observations executed by the Short-term scheduler according to the
process described.}}
\label{fig:schedST}
\end{center}
\end{figure}

First, the algorithm removes all objects whose assigned observation period ended
before the current time from the mid-term plan and selects the next target.
For this target, the code computes the slew time of the telescope. The target
observation obtained for the mid-term plan is adapted (i.e., advanced or delayed)
according to the current time, computed slew time, and integration time. This observation
is only selected if it fulfills the hard constraint requirements until the end of
the observation. Otherwise, it is discarded and the gap between the current time
and the start of the next observation in the mid-term plan is filled.
The filling process sorts all the observations that (1) are not already in the
mid-term plan, (2) fulfill the hard-constraints during the entire window, and (3)
can be completed in the available time. This ranking is done according to several
criteria described below. Finally, the first observation in the sorted list is
selected as the next observation. The filling process is repeated until the gap
is filled or there are not target observations left.
Each target selected by the Short-term scheduler is sent to the
Instrument Control System, and the information on the success of the observation is
stored in the database for use in subsequent scheduler runs. Besides, the
sorted list of objects can be provided to the operator for override in case
of need. Fig.~\ref{fig:schedST} shows an example of the most
common situation where the Short-term scheduler needs to repair the mid-term
plan to select the next target to be observed.

%\paragraph{Observation list sorting.}\label{s:st_sched:list}
The ranking of the targets is key in the process of repairing the mid-term plan
by filling a gap between the last executed observation and the next observations
recommended by the mid-term plan. This ranking is based on astronomical heuristics,
and the targets are sorted according to the first rule, then the second rule, and
so on. The defined rules are:
\begin{enumerate}
 \item The number of times that the target is observed during the current night
 (smallest to largest).
 \item The target is not in the remaining mid-term plan.
 \item The priority of the target (largest to smallest).
 \item The number of times that the target has been observed in the survey
 (smallest to largest).
 \item The proximity to meridian crossing according to Eq.~\ref{eq:midFitness1}
 (largest to smallest).
\end{enumerate}

The main idea of this process is to fill the gaps with interesting objects at the
current time, according to the times that they have been observed, their priority, and
the proximity to meridian crossing. Rule 2 is key to fill gaps without affecting
excessively the mid-term plan, which has been globally optimized.

\section{CAST scheduling efficiency}
\label{sec:efficiency}

In general, CAST is focused on optimizing the three soft constraints described
in Sect.~\ref{sec:CAST}: observing time, observation deviation and observation sequence. The first two constraints
maximize the use of the telescope and the instrument while the last one is
included as an optional condition to increase the scientific return.

In this section we run a set of simulations of the CARMENES survey with the aim
of analysing the efficiency of CAST in the use of resources. For a quantitative
analysis, we have defined different metrics related to the use of the telescope
and the instrument. In particular, we have computed the fraction of targets
that are planned by the scheduler, the total number of observations, the
fraction of available time that the telescope is operating, the fraction of
time during which the instrument is performing science observations, and the
fraction of overhead time. In Sect.~\ref{sec:science} we describe the impact
of the observing plans optimized by CAST on the CARMENES scientific results.

\subsection{CARMENES configuration} \label{subsec:configuration}
To be as close as possible to the real survey, we have adopted the following
procedure to run CAST. The Long-term scheduler has a scope of six months,
and is executed every three months during the survey. The Mid-term
scheduler is executed every day taking into account the observations acquired
during the previous night. The Short-term scheduler is executed ``on the fly''
each time a new observation is required and takes into account the mid-term
plan, the observations already carried out during the night, and any variation
on the weather or instrument conditions.

In terms of parametrization, the Long-term and Mid-term schedulers have
several parameters related to the GA as explained before. Table~\ref{tab:parameterValuesGA}
summarizes the parameter configuration used in the experiments done, { which
are related to the number of generations of the evolutionary algorithm,
the number of elements in the initial set of solutions and in the subsequent
generations and the probabilities of selection, crossing, and mutation.}

\begin{table}[!t]
\begin{center}
\begin{footnotesize}
\caption{Genetic algorithm parameter configuration.}
\label{tab:parameterValuesGA}
\begin{tabular}{l l}
\hline
\hline
\noalign{\smallskip} 
Parameter & Value\\
\noalign{\smallskip} 
\hline
\noalign{\smallskip} 
Long-term generations & 1000 \\
Mid-term generations & 1000 \\
$N_I$ & 50 \\
$N_P$ & 100 \\
$p_s$ & 0.4 \\
$p_c$ & 0.9\\
%$w$ (only for Long-term) & 0.8 \\
$p_\mu$ & $1/\overline{\overline{I}}$\\
\noalign{\smallskip} 
\hline
\end{tabular}
\end{footnotesize}
\end{center}
\end{table}

\subsection{Results}
\label{subsec:results_sched}

Because GAs are stochastic methods, CAST is executed 50 times with different
random seeds with the aim of avoiding any bias in the results due to
convergence to local minima. Hereafter, each of these executions is referred to
as a trial. Table~\ref{tab:effResults1} summarizes the parameters of the
simulations and the results of the metrics used to evaluate the efficiency of
CAST.

From the the {simulated} weather statistics, the average usable time for
observations at Calar Alto is around 60\,\% of the total night time.
{Observatory statistics
actually indicate that $\sim 70$\,\% of the nights are useful according to
meteorological variables \citep{Sanchez07,Sanchez08}. However, as explained
before we have considered and additional up to 20\,\% of lost time due to
cloudiness or technical issues.}
Thus, in three years, there are about 6400 hours
during which observations can be scheduled. The
results from our simulations presented in Table~\ref{tab:effResults1} show that
all required targets are always planned. Around { 21000} observations are
scheduled, occupying $\sim$99\,\% of the good weather time. The break-down of
this time indicates that { 84\,\%} corresponds to the telescope collecting photons
and { 16\,\%} is spent during slews to new positions. This means that the
instrument is collecting photons during approximately { 1700} hours per year. Our
simulations show that CAST can optimize the observing time of the telescope by
selecting the best targets to observe according to environmental conditions at
each time. In our simulations, the average integration time of the targets is
around { 14 minutes} and we can obtain about { 3.2} observations per hour
of working time.

\begin{figure*}[!t]
    \center
   \includegraphics[width=1.0\textwidth]{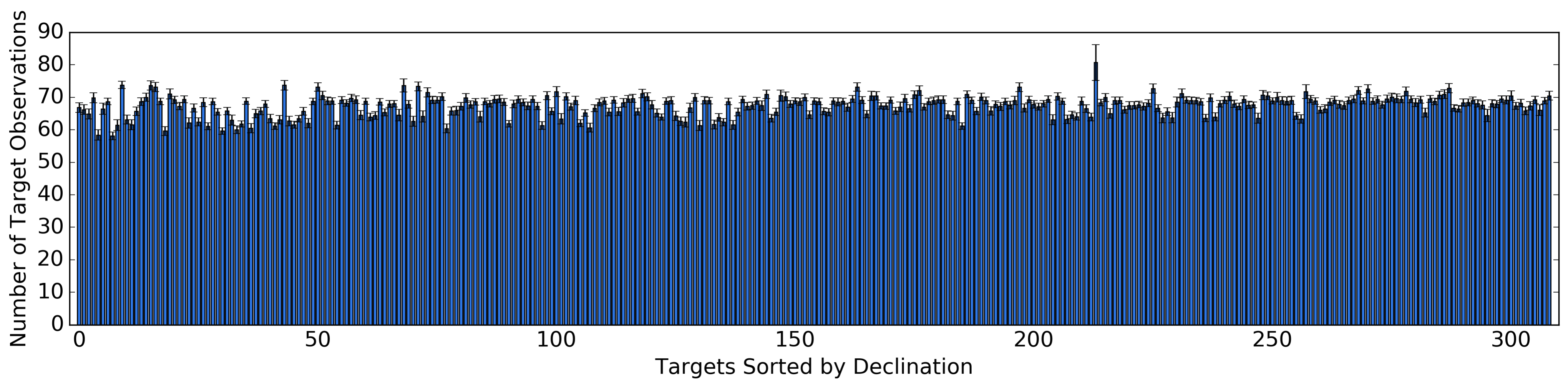}
    %\vspace*{0.2cm}
    \caption{Number of observations scheduled for each target. The horizontal
axis represents the identifier of each one of the 309 targets used in our
simulations sorted by increasing declination. Declination ranges from $-21$ to
83\,deg in our sample. The bars show the average and standard deviation values
of the 50 executions.}
    \label{fig:targetValidation}
\end{figure*}

Additionally, it is important to distribute equitably the observations of all
the targets. This is the second soft constraint that CAST must optimize.
Fig.~\ref{fig:targetValidation} shows the number of times that each target is
observed. On average, each M dwarf in the sample is observed { $\sim$67} times
during the three-year survey. The standard deviation of the number of
observations between targets is { 3}, which indicates that the resources are
reasonably distributed among the different targets. Only a few of them have a
number of observations significantly below the average, but this is due to
their very limited visibility during the year (e.g., low declination and
faintness). This means that all targets will have a high number of
observations and, if necessary, this number can still be increased if some of
the targets are discarded during the initial sample clean-up (very fast rotators,
active stars, spectroscopic binaries that passed our filters, etc.). 
On the other hand, the simulations with CAST could also
help in the optimization of the sample since any target for which a minimum
number of observations is not reached can be rejected, or upper limits to the
number of observations per target can be set.

As already mentioned, CAST includes a soft constraint to define an observing
scheme that maximizes planet detectability with the least number of
observations. For the present work we have assumed a maximum of one observation
per night for any target, but a well-sampled window function of the
observations may improve the detectability of signals within a range of
interesting periods. This can be quantified by computing a phase dispersion
parameter given a set of previous observations for the periods of interest of
each target and then estimating the next best observation that optimizes such
parameter \citep[see, e.g.,][]{Freedman94}. These techniques to optimize the
periodogram window function can only be applied once several observations per
target are obtained.

\begin{table}[!t]
\caption{CAST parameters and mean values of the metrics.}
\vspace*{-0.35cm}
\begin{center}
\label{tab:effResults1}
\begin{footnotesize}
\begin{tabular}{p{4.2cm}p{3cm}}
\hline
\hline
\noalign{\smallskip} 
\multicolumn{2}{l}{CAST parameters {(a)}} \\
\noalign{\smallskip} 
\hline
\noalign{\smallskip} 
Days planned & 1096 \\
Total targets & 309 \\
Total observable time &  10703.05\,h \\
Unfavorable weather time &  4300.23$\pm$88\,h \\
Available time for observations {(b,f)}&  59.82$\pm$0.82\,\% (6402.81\,h)\\ % \newline (3197\,h 56\,m $\pm$77\,h 14\,m)\\
Execution Time of CAST & 23.85$\pm$0.11\,h \\
%\hline
\noalign{\smallskip} 
\hline
\noalign{\smallskip} 
\multicolumn{2}{l}{Metrics {(a)}}\\
\noalign{\smallskip} 
\hline
\noalign{\smallskip} 
Planned targets {(c)} & 100$\pm$0\,\% \\
Observations done & 20827$\pm$293\\
Working time ({d,f}) & 99.05$\pm$0.06\,\% (6342.03\,h) \\
Tracking time ({e,f}) & 84.18$\pm$0.03\,\% (5338.77\,h) \\
Overhead time ({e,f}) & 15.82$\pm$0.03\,\% (1003.22\,h) \\
\noalign{\smallskip} 
\hline
\end{tabular}
%\tablefoot{
\begin{tiny}
 (a){The uncertainties are computed as the standard deviation of 50
random trials.}\\
(b){Fraction of time available for observations, excluding bad
weather time.}\\
(c){Fraction of the targets that are planned.}\\
(d){Fraction of time with scheduled telescope operations with
respect to the available time for observations.}\\
(e){Fraction of time with respect to the working time.}\\
(f){The number of hours is indicated in parentheses.}
\end{tiny}
%}
\end{footnotesize}
\end{center}
\end{table}

\subsection{Analysis of the computational performance}
\label{subsec:pc_performance}

The computing time of the three scheduler steps with the 309 selected targets
is presented in Table~\ref{tab:computationalCost}. CAST has been developed in
C$++$ language and the experiments have been executed using only one processor
of the main computer where CAST runs in Calar Alto (Dell PowerEdge R420 rack server
with two Intel Xeon CPUs E5-2430 v2 with six cores at 2.50\,GHz and 16\,GB of
RAM). The main restriction in the scheduling process is the response of the
Short-term scheduler. As already explained, this is used to select the next
best survey target after the end of each observation by taking into account the
environmental conditions. It is required that its execution takes less than 5
seconds to prevent telescope idle time during the calculation. This requirement
is fulfilled as the Short-term scheduler selects a new observation in about 56
milliseconds. The other schedulers can be executed during the day before the
observations, and they do not have any time restriction. Nevertheless, they
produce results in a reasonable time: about { 30 minutes} and $<1$ minute for
the Long-term and Mid-term schedulers, respectively.

\begin{table}[!t]
\caption{Average computing times of the three scheduler strategies in the 50
trials.}
\vspace*{-0.35cm}
\begin{center}
\label{tab:computationalCost}
\begin{footnotesize}
\begin{tabular}{lr}
\hline
\hline
\noalign{\smallskip} 
Scheduler & Execution Time (s) \\
\noalign{\smallskip} 
\hline
\noalign{\smallskip} 
Long-term scheduler  & 1815.166 \\
Mid-term scheduler   & 58.347 \\
Short-term scheduler & 0.056 \\
\noalign{\smallskip} 
\hline
\end{tabular}
\end{footnotesize}
\end{center}
\end{table}

\section{CAST in real operation}
\label{sec:real_operation}
CAST is in real operation since September 2016 for the CARMENES survey. The
used version includes some additional functionality that is specific to the
observing program and that is intended to adapt its performance to the actual
working conditions. We list here some of these additional specific features
that extend beyond the general design described above:
\begin{itemize}
\item During the survey it is possible to change 
the priority of an object to increase the chances of it being selected by CAST.
\item It is possible to disable an object to make it unavailable for CAST.
\item CAST can include the observations of telluric standard
stars during the nautical twilight in the planning.
\item Bright stars can also be planned during nautical twilight updating the
\emph{Night} constraint of the affected objects with the desired Sun altitude.
\item A parameter is available to select the maximum number of times that a
target can be observed during one night.
\item It is possible to assign a cadence of observations to each target
to sample different periods (from days to months).
\item Interesting or standard stars can be labeled as mandatory. This implies
that they must be observed every night while they have this flag active.
This constraint is also considered by the Mid-term scheduler in two steps:
\emph{1)} the Mid-term scheduler is executed only with the mandatory objects,
following the same defined objective functions, and
\emph{2)} the Mid-term scheduler is executed a second time with the remaining
objects but blocking window times where the mandatory objects have been
planned. The final mid-term plan is the merging of both plans.
\item The quality of the night according to transparency conditions can be considered.
This quality can be modified during the night by the operators. There are
six keywords indicating different night quality levels: excellent, good, fair,
poor, very poor and bad. When an observation is required with a night quality
that is not excellent, the mid-term plan is not considered and the short-term
scheduler is in charge of selecting the next observation. In this situation,
the rules specified in Sect. \ref{s:st_sched} are modified as follows:
\begin{enumerate}
 \item The proximity to meridian crossing according to Eq.~\ref{eq:midFitness1}
(largest to smallest) weigthed with $R_{\rm q}$ following equation
\begin{tiny}
\begin{align}
\label{eq:badnight}
\begin{split}
&R_{\rm q}(t,q)=\left(\frac{t_{\rm magnitude}}{4.2}\right)^{(q-5)}\ ,
\end{split}
\end{align}
\end{tiny}
\item[] which relates the magnitude of an object with the quality of the night in order
to increase the ranking of bright objects. In Eq.~\ref{eq:badnight},
$t_{\rm magnitude}$ is the $J$-band magnitude of target $t$, and $q$ is the
quality of the night that goes from 4 (good) to 0 (bad).
 \item The number of times that the target is observed in the current night
(smallest to largest).
 \item The priority of the target (largest to smallest).
 \item The number of times that the target has been observed in the survey
(smallest to largest).
\end{enumerate}
\end{itemize}

\section{The impact of efficient scheduling on the scientific return}
\label{sec:science}

\subsection{Exoplanet yield simulation}
\label{subsec:planets_simulation}

In this section we analyze the impact of our scheduler algorithms in the
scientific results of the CARMENES survey. We estimate how many planets would
be detected in a radial-velocity survey following our planning scheme. As
described in the introduction, the CARMENES survey has been designed to
discover a statistically significant sample of exoplanets orbiting M dwarfs
\citep{Quirrenbach14}. We have shown that CAST can optimize the survey to
obtain over { 67} observations per target distributed during the three years of
operations. To evaluate the exoplanet yield, we have performed a number of
simulations that are described below.

We simulated 100 hypothetical planet scenarios using the stellar properties
available for our 309 potential targets in the CARMENES database: right
ascension, declination, magnitude (which are input parameters of the scheduling
algorithms) and spectral type.
We used mass-luminosity relations to estimate the mass
of each target. 2MASS $K_{s}$-mag and distance are taken from Carmencita 
\citep{Caballero13a,AlonsoFloriano15} to derive absolute $K$-band magnitudes,
and masses are computed using calibrations from \citep{Delfosse2000}.

The properties of the simulated exoplanets were estimated according to the
currently available exoplanet statistics. We used functional fits instead of
the tables given in papers for convenience and to approximately reproduce the
expected behaviour of the probability distributions with mass and period.
However, these distributions should be taken with some caution given the
uncertainties of planet ratio tables. The specific details for the different
parameters were as follows:
\begin{itemize}
\item We used the mass distribution from \citep{Mayor11}, {which considers
planets orbiting FGK type stars. Planet rates for M-dwarfs are also provided by
\citep{Bonfils13b}, but they are based on a smaller sample of planets, thus with
large uncertainties or only upper limits.} However, since we are
interested in M dwarfs, we scaled the planet rates of the largest mass
bins in \citep{Mayor11} following the statistics in \citep{Dressing13}.  As a
smooth approximation to the binned statistics, for our simulations we fitted an
exponential function to the resulting planet rates. We considered planets with
masses between 1~M$_{\oplus}$ and 1000~M$_{\oplus}$, by extrapolating the fit
below 3~M$_{\oplus}$. We set a lower limit to the planet mass at
1~M$_{\oplus}$ because, although we expect that sub-Earth mass planets are very
abundant, their statistical distribution is not sufficiently well constrained
to enter our simulations. Furthermore, such low-mass planets are unlikely to
be detectable by CARMENES.
\item To assign the orbital periods, we proceeded in the same way by using the
planet rates for giant and large Neptune-like planets given in \citep{Fressin13}
for simulated planets with masses above 30~M$_{\oplus}$, and those given by
\citep{Dressing15} for smaller planets. In this case, we fitted a second order
polynomial in a log-log scale to take into account the lower probability of
planets on short periods. In our simulations we assumed 0.5~days as the
minimum orbital period of planets.
\item The eccentricity was assumed to follow the distribution presented in
\citep{Kipping13}.
\item The inclination was taken as uniformly distributed in $\sin i$, taking
into account all the possible orientations of the orbital plane.
\item The argument of the periastron was assumed to follow a uniform
distribution between 0 and 360\,deg.
\item The multiplicity of exoplanets systems was adopted to follow the Kepler
Objects of Interest statistics for 1135 stars (58.9\,\% of single planet systems, 26.5\,\% of
double planet systems, 8.6\,\% triple, 4.3\,\% quadruple, 1.3\,\% quintuple, 0.2\,\%
sextuple, and 0.2\,\% heptuple). These multiplicity rates
slightly underestimate the fraction of planets per star in our simulations because of
geometrical effects (narrower orbital inclination range for transits as the number of
detected planets increases). However, this should not affect the conclusions of our
analysis. Assuming that all stars have planets, our statistics of multiple
systems predicts about 1.6 planets per star, close to the range of values given by
\citep{Gaidos2014} and \citep{Dressing15}. In the case of multiple systems, for our
simulations we imposed a conservative minimum ratio between planet
periods of 1.3 to ensure realistic configurations in terms of orbital stability.
\end{itemize}

More details on the functional relationships and assumptions are provided
in \citep{Perger2017}.

Our method to validate the scientific efficiency of the CARMENES survey using
CAST is described as follows:
\begin{enumerate}
 \item Firstly, we generated a planet scenario $s$ as described above. For
statistical significance, we simulated 100 different scenarios ($N_S$), i.e.,
100 different planet realizations, for each of the 309 targets ($N_T$) in the
scheduling plan described in Sect.~\ref{sec:efficiency}. {As a consistency
check, we compared the statistics of our planet scenarios with those in
\citep{Bonfils13b}, finding that planet ratios are in agreement within the error bars
for the planetary mass and orbital period bins relevant for our analysis.}
 \item For each exoplanet scenario $s$, we applied the 50 scheduling plans
($N_R$) obtained {in Sect.~\ref{sec:efficiency}}. Each scheduling trial $r$ comprises
three years of observations.
 \item For each target, we calculated the Keplerian radial-velocity curve that
would be obtained according to the exoplanet scenario $s$ and the observation
dates of the scheduling trial $r$. In the case of multiple systems,
we added the contribution of each planet linearly. We also included a
white noise contribution to take into account the measurement uncertainty. In general,
a precision of 1~\ms\ was considered for all targets, except for targets
fainter than $J$=8~mag, for which the photon noise was assumed to scale with
the magnitude from the fiducial value of 1~\ms\ at $J$=8\,mag in 875\,s.
 \item In order to identify the planets that would be detected with CARMENES,
we obtained the periodogram of the observations for each target following
\citep{Zechmeister09}. We labeled a planet as identified when the peak of the
periodogram corresponded to the period of the planet simulated and the false
alarm probability (FAP) was below the 0.1\% threshold. To account for possible
multiplanet systems, we subsequently computed the periodogram of the residuals
subtracting the periods detected sequentially until no further significant
periods were found.
\item Significant signals that did not correspond to any of the simulated
planets were labeled as false positives.
 \item Finally, we computed the average number of detected planets and false
positive detections over all planet scenarios and scheduling trials to evaluate
the efficiency on discovering planets of the simulated survey and their properties.
\end{enumerate}

\subsection{Results}
\label{subsec:results_planets}

We tested 100 different random exoplanet scenarios and obtained the average
number of planet detections for 50 scheduling trials, thus adding to a total of 5000
simulations. In particular, we computed the average number of detected planets,
the average number of detected planetary systems (considering that a system is
detected if all its planets are found), and the average number of false
positive detections. In order to assess the scientific return of the simulated
survey, we analysed the number of planets that we expected to detect and their
properties.

\begin{figure}[!t]
    \center
    \includegraphics[width=0.5\textwidth]{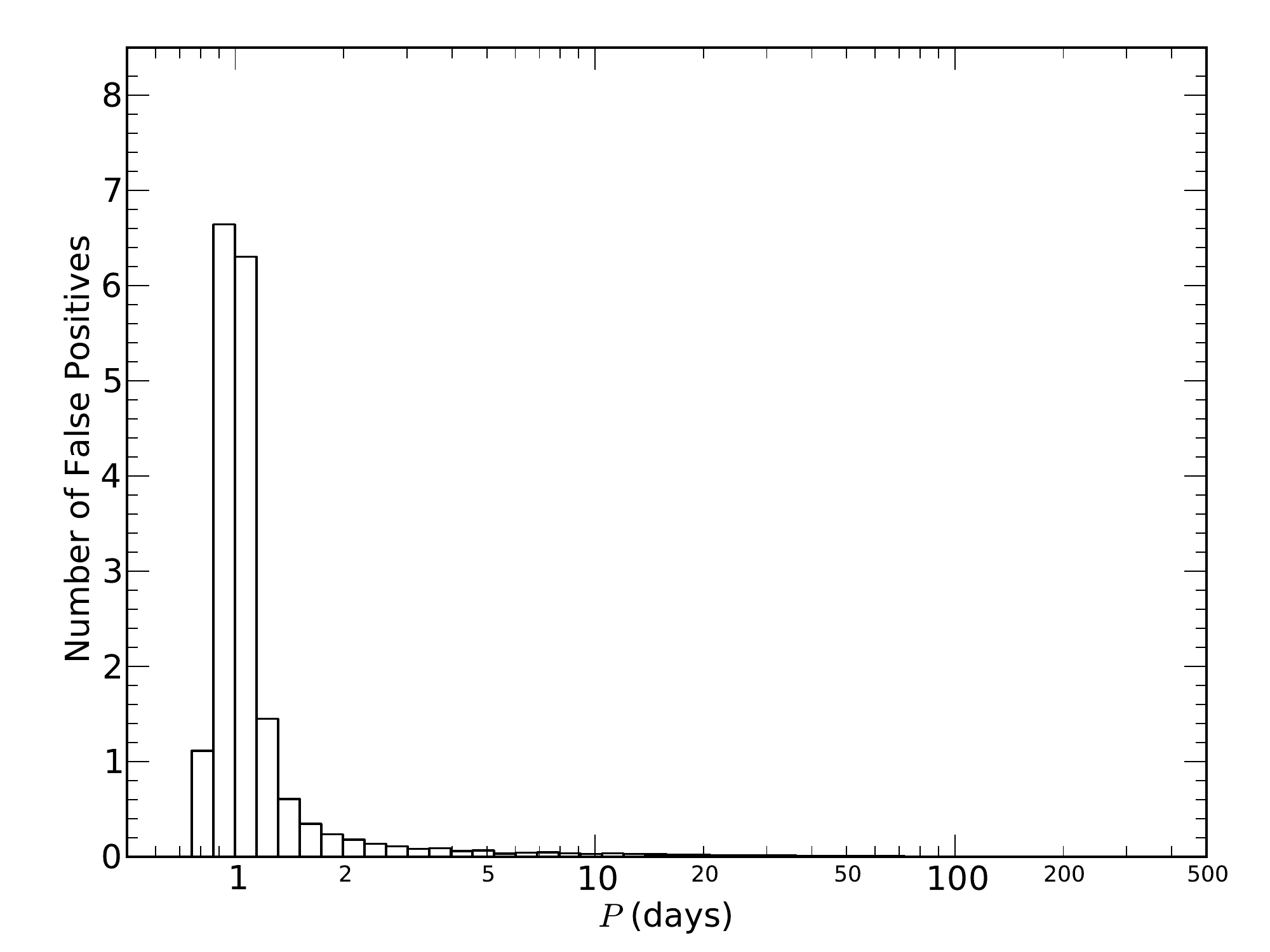}
    \caption{Histogram of the mean values of false positive detections as a function
of period. The mean values are computed for the set of 5000 simulations (50 planning
and 100 scenarios).}
    \label{fig:falsepositives}
\end{figure}

Each of the scenarios contained a different number of planets orbiting the
stars selected from the Carmencita database according to the adopted planet
distribution functions. On average, a total of { 505$^{+16}_{-17}$} planets were
generated for each scenario. In the absence of any noise source other than photon
noise, { 118$^{+9}_{-9}$} planets were detected following the usual periodogram
analyses (correct period and FAP$<$0.1\%), with { 3.6$^{+2.4}_{-1.6}$} planets that
could be transiting. With respect to multi-planetary systems, for a total of
{ 3.9$^{+1.1}_{-1.9}$} stars all their planets were found in the periodogram. Finally,
there were { 17.9$^{+4.8}_{-4.9}$} false-positive detections, representing a
false-positive rate of about 13\% on average. Fig.~\ref{fig:falsepositives} shows the
distribution of false positives as a function of period. Most of these
false positives correspond to the obvious 1-day sampling alias and could be readily identified
and discarded. The few cases of false positives with long periods were due to
residual signals from eccentric planets. All this means that the planet detection method
is sensitive enough to identify planets without suffering excessively from incorrect
detections.

\begin{figure}[t!]
    \center
    \includegraphics[width=0.43\textwidth]{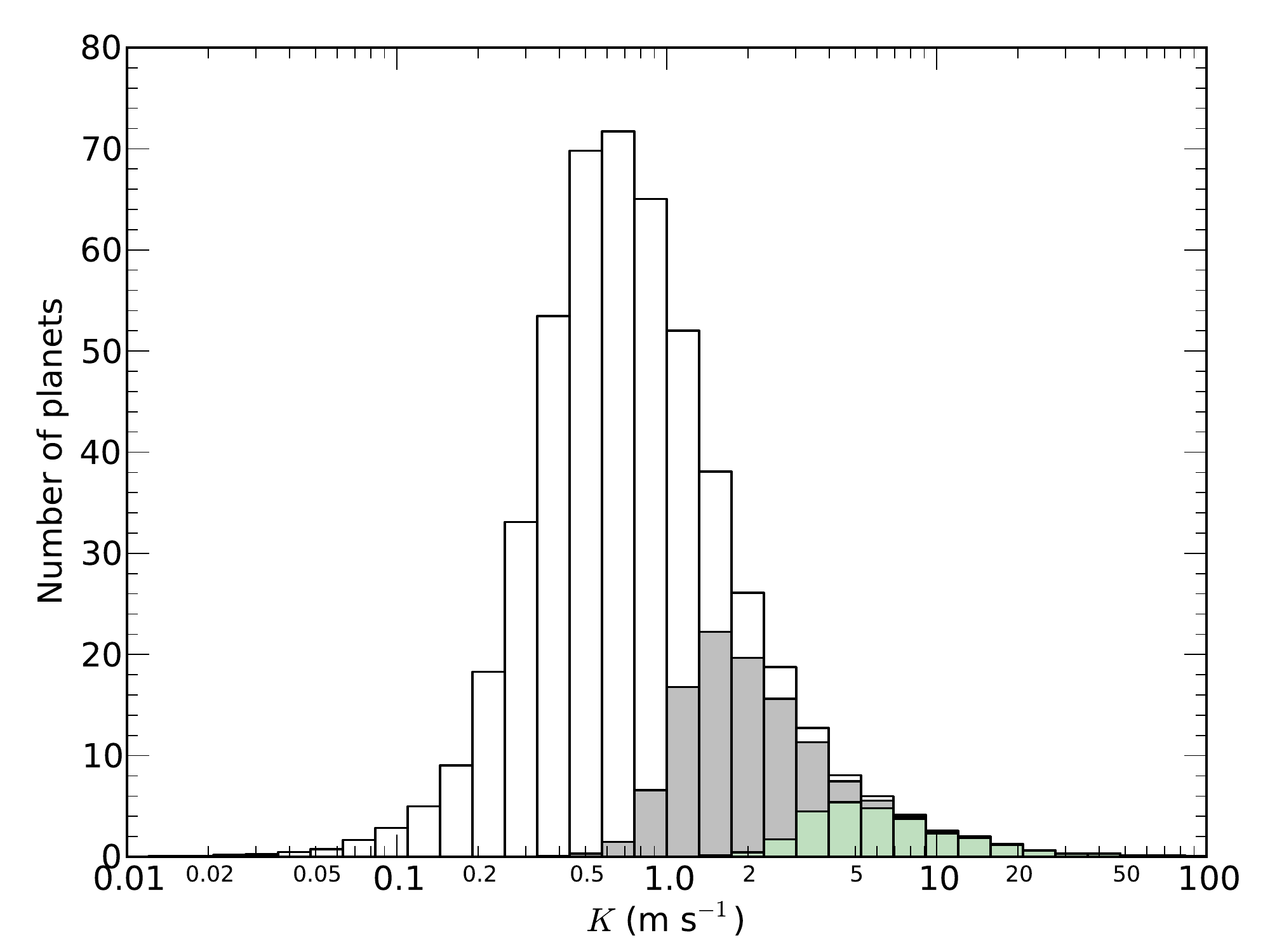}
    \includegraphics[width=0.43\textwidth]{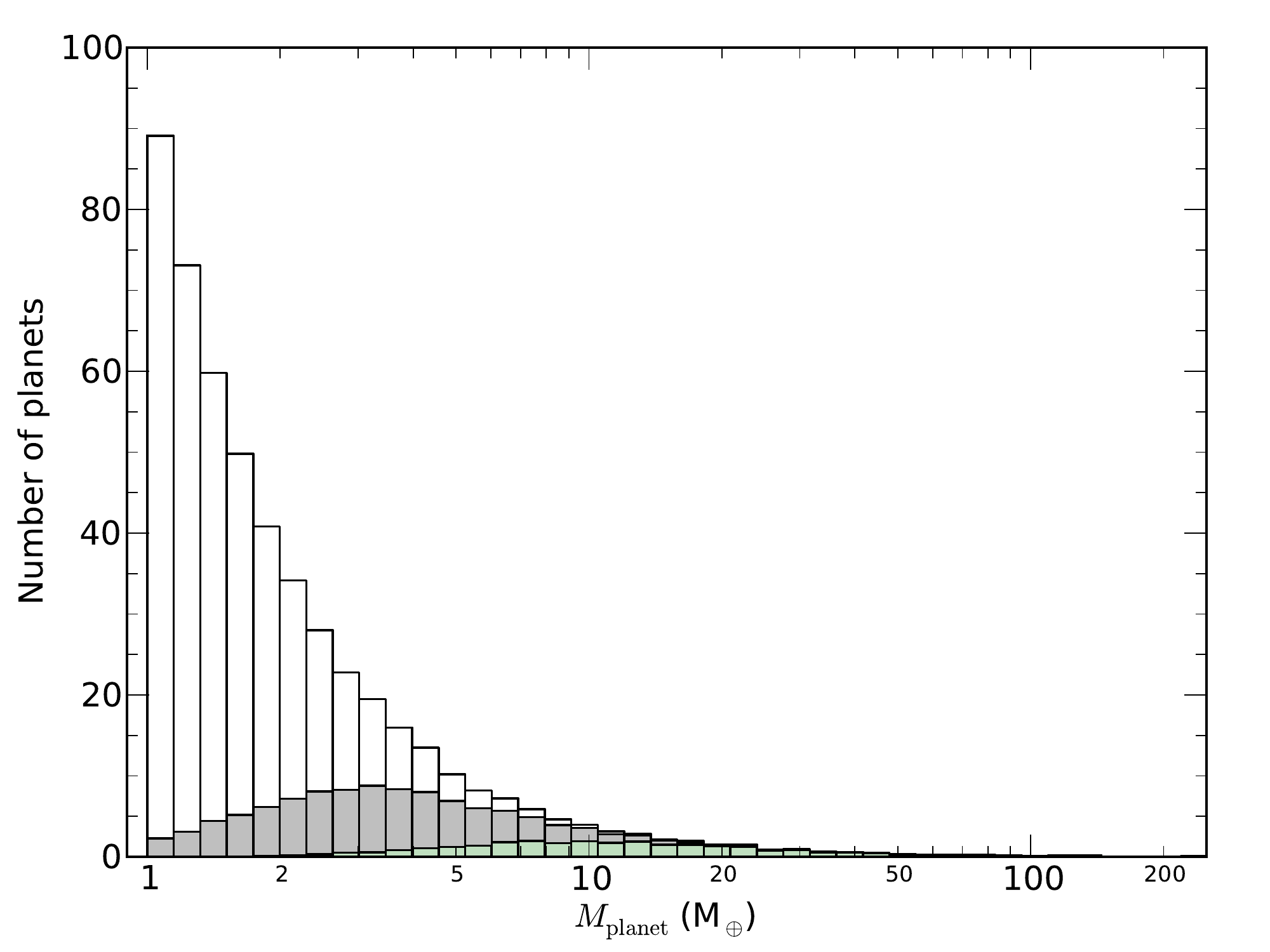}
    \includegraphics[width=0.43\textwidth]{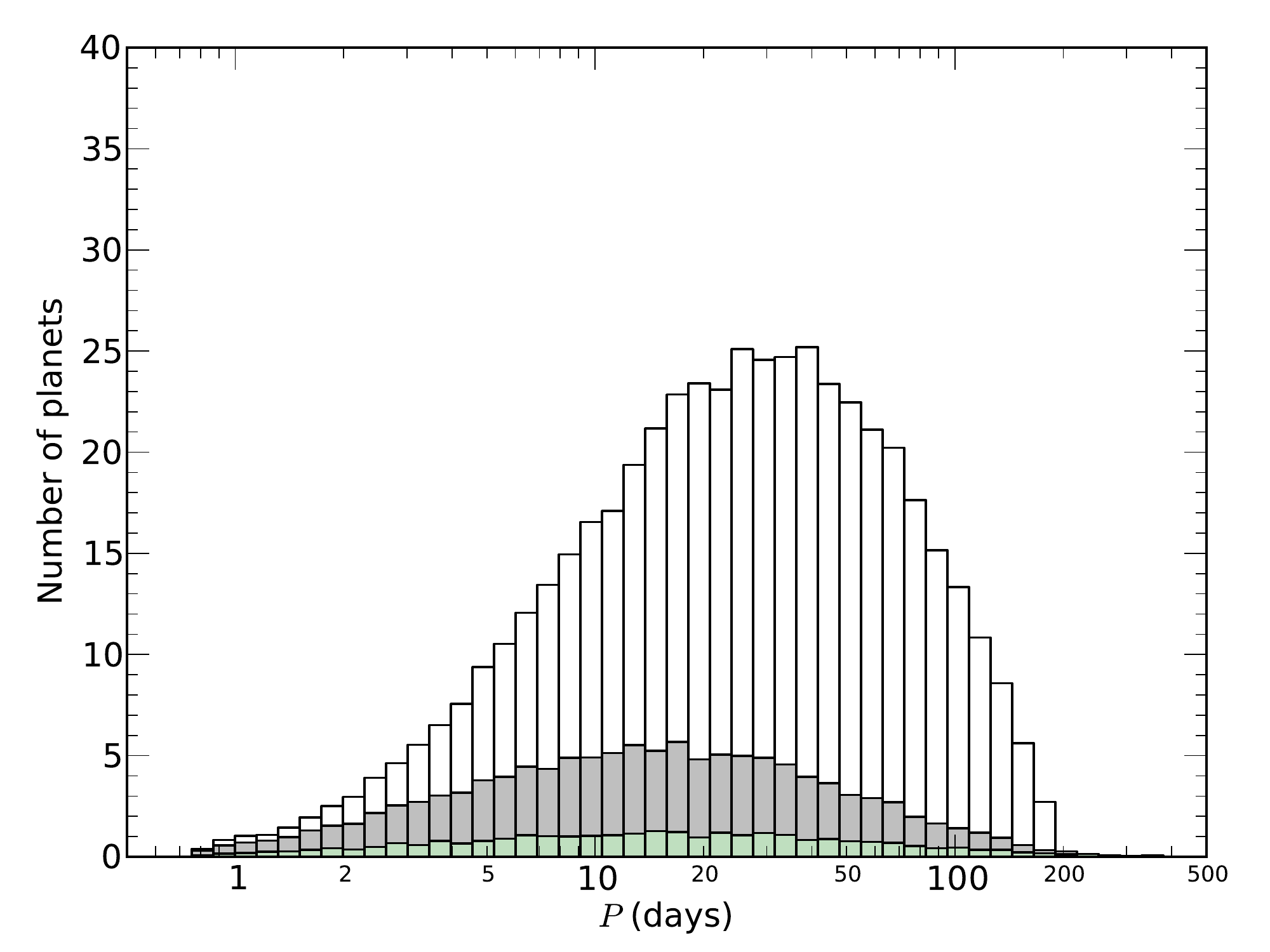}
    %\vspace*{0.2cm}
    \caption{Histogram of the radial-velocity semi-amplitude (top), planet mass
(middle) and orbital period (bottom) of the generated planets in our
simulations. In all panels, open bars show the average number of simulated
planets for the 100 scenarios and 50 plannings for reference. The number of
detected planets assuming 1\,\ms\, pure photon noise is shown as coloured bars.
Detections with 3\,\ms\ stellar activity jitter added to photon noise are shown
in green.}
    \label{fig:histograms}
\end{figure}

In percentage, { 23.4$^{+1.9}_{-2.0}$\%} of the generated planets were detected in our
simulated survey, while the average number of multiplanetary systems found was
3.1$^{+1.8}_{-1.5}$\%. To understand this efficiency, one must keep in mind that a
significant fraction of the simulated planets lay under the detection limit of
the CARMENES instrument (in terms of radial-velocity amplitude). The top panel
in Fig.~\ref{fig:histograms} shows the histogram of the planet radial-velocity
semi-amplitudes. The average number of generated planets and those that are
detected are shown as open and shaded bars respectively. The steep decrease of
the overall distribution at low amplitudes is not a physical effect, but a
consequence of the limit in planet mass at 1\,M$_{\oplus}$. We find that
exoplanet signals with semi-amplitudes exceeding { 0.80~\ms\ can be detected
with 89\,\% probability according to our simulations.}

An effect to consider in M-dwarf star surveys is the impact of magnetic activity
causing so-called radial-velocity jitter \citep{Martin06,Prato08}. In the
CARMENES sample that we have used in our simulations, 30\% of the M dwarfs
show the H$\alpha$ line in emission (see Fig.~\ref{fig:targets}), which is a
signpost of moderate to high activity levels (Jeffers et al., submitted).
To provide a rough assessment of the potential effect of activity in the CARMENES survey,
we run a set of simulations by considering an additional white noise term added
in quadrature to the measurement uncertainty simulated.
{Although radial velocity jitter could be larger towards
more active later spectral types, as a simple approach we have taken the mean noise
value of 3\,\ms\ reported in \citep{Perger2017}. However, we recall that in our
simulations the radial velocity uncertainty is also larger for these
stars because they are generally fainter.}
 The results are also shown in Fig.~\ref{fig:histograms} as
green bars. As expected, the radial-velocity threshold for detecting planets
increases and the number of detections is reduced to about { 28 planets.
This is a simple approach assuming non-correlated stellar jitter. However, 
a thorough analysis of radial velocities using the
wide wavelength interval covered by CARMENES (0.52~$\mu$m)--1.71~$\mu$m)) helps to disentangle activity noise
from exoplanet signals and improve the number of detections.}

Fig.~\ref{fig:histograms} also shows histograms for the planet mass (middle)
and orbital period (bottom) of the generated and detected exoplanets (both with
and without activity jitter). As expected, long-period planets are more
difficult to find because the reflex radial velocities of the host stars have
lower amplitudes. On the other hand, almost all planets with masses above
5\,M$_\oplus$ are above the detection threshold.

Our simulations show that the use of CAST for the planning of observations
guarantees that a high percentage of the planets expected to be discovered by
CARMENES could be identified. Table~\ref{tab:activity} summarizes the number
and fraction of detected planets given all simulated systems and also 
considering the 1\,\ms\ threshold. The results taking into account an
additional 3\,\ms\ activity jitter are also shown for comparison, but this is
a worst-case scenario, as explained above.

\begin{table*}[t]
\caption{Mean values of the 50 executions of 100 planet scenarios assuming
different levels of stellar intrinsic radial-velocity jitter ($\sigma_{\rm
activity}$).}
\vspace*{-0.35cm}
\begin{center}
\label{tab:activity}
\begin{footnotesize}
\begin{tabular}{p{0.8cm}p{1.3cm}p{1.3cm}p{1.3cm}p{1.3cm}p{1.5cm}p{1.7cm}p{1.7cm}}
\hline
\hline
\noalign{\smallskip} 
$\sigma_{\rm activity}$ [m$\thinspace$s$^{-1}$] & Generated planets & Detected planets & Generated systems & Detected systems & False \mbox{Positives} &  Detected \mbox{planets (\%)} & Detected \mbox{systems (\%)} \\
\noalign{\smallskip} 
\hline
\noalign{\smallskip} 
%All planets \\
\multicolumn{8}{l}{All planets} \\
\noalign{\smallskip} 
\hline
\noalign{\smallskip} 
 \mbox{0}   & \mbox{505$^{+16}_{-17}$} & \mbox{118$^{+9}_{-9}$} & \mbox{126$^{+11}_{-7}$} & \mbox{3.9$^{+1.1}_{-1.9}$} & \mbox{17.9$^{+4.1}_{-4.9}$} & \mbox{23.4$^{+1.9}_{-2.0}$} & \mbox{3.1$^{+1.8}_{-1.5}$}   \\
 \mbox{3}   & \mbox{505$^{+16}_{-17}$} & \mbox{\phantom028$^{+5}_{-6}$}   & \mbox{126$^{+11}_{-7}$} & \mbox{0.2$^{+2.8}_{-0.2}$} & \mbox{\phantom04.8$^{+2.2}_{-2.8}$}  & \mbox{\phantom05.5$^{+1.2}_{-0.9}$}  & \mbox{\phantom00.2$^{+2.5}_{-0.2}$}   \\
\noalign{\smallskip} 
\hline
\noalign{\smallskip} 
%Planets with RV > 1 m/s \\
\multicolumn{8}{l}{Planets with $K \ge 1$~\ms} \\
\noalign{\smallskip} 
\hline
\noalign{\smallskip} 
 \mbox{0}   &  \mbox{174$^{+10}_{-12}$} & \mbox{110$^{+7}_{-12}$} & \mbox{26$^{+5}_{-3}$} & \mbox{6.4$^{+2.6}_{-2.4}$} & \mbox{17.9$^{+4.1}_{-4.9}$} & \mbox{63.3$^{+3.8}_{-4.0}$} & \mbox{24.2$^{+9.1}_{-8.1}$} \\
 \mbox{3}   &  \mbox{174$^{+10}_{-12}$} & \mbox{\phantom028$^{+5}_{-6}$} & \mbox{26$^{+5}_{-3}$} & \mbox{0.3$^{+5.7}_{-0.3}$} &  \mbox{\phantom04.8$^{+2.2}_{-2.8}$} & \mbox{16.0$^{+3.2}_{-2.6}$} & \mbox{\phantom01.3$^{+22.7}_{-1.3}$}   \\
 \noalign{\smallskip} 
\hline
\end{tabular}
\end{footnotesize}
\end{center}
\end{table*}

\begin{figure}[!t]
    \center
   \includegraphics[width=0.43\textwidth]{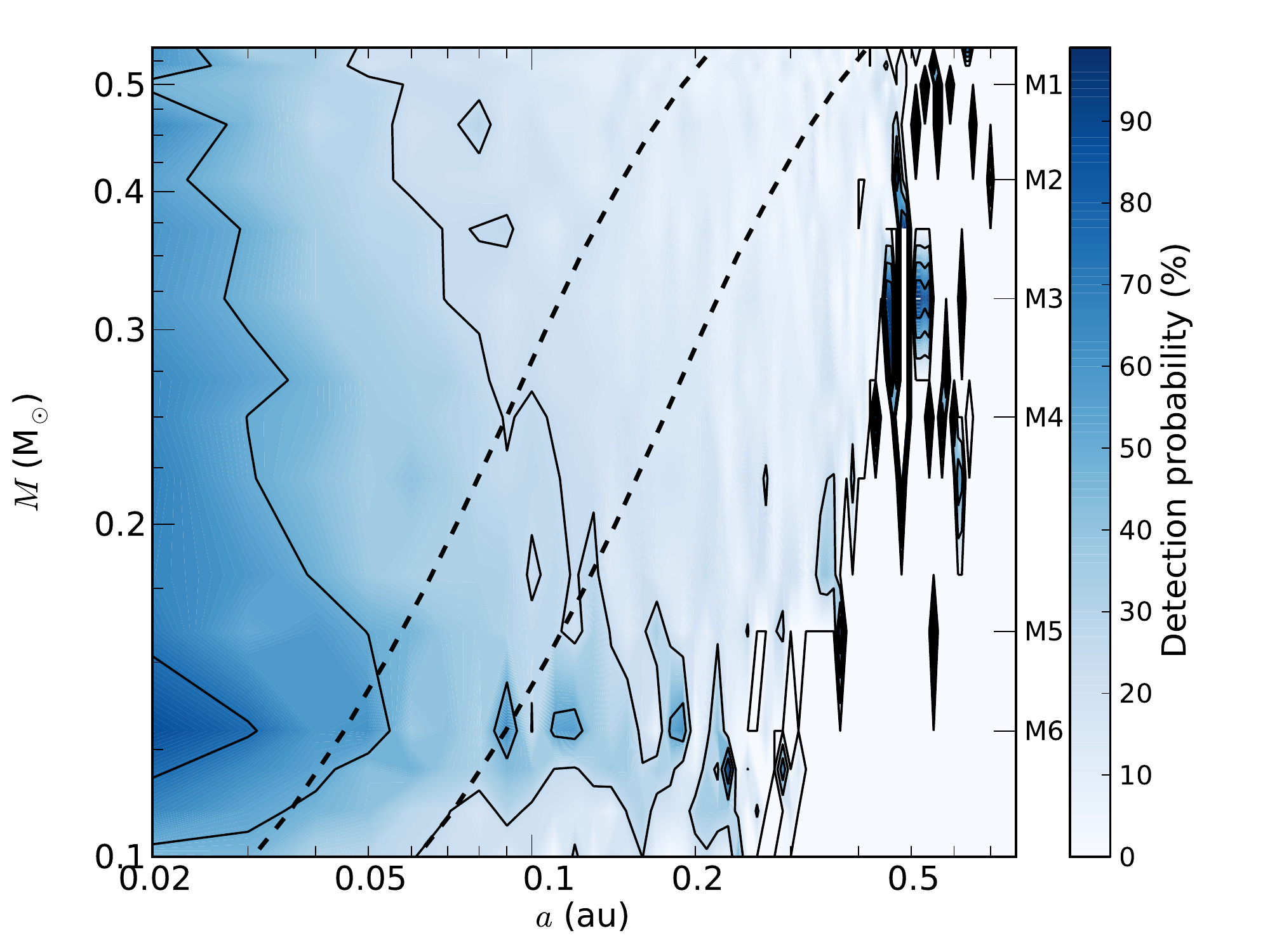}
%   \includegraphics[width=0.48\textwidth]{./HZplot_rv3.pdf}
    %\vspace*{0.2cm}
    \caption{{Probability of detection of the planets generated in CARMENES
survey simulations in a semimajor axis vs. host star mass diagram. Dashed lines
indicate the inner and outer habitable zone limits for 1~$M_{\oplus}$ planets
given in \citep{Kopparapu14}. Contour lines for 25, 50 and 75\% detection probability
are shown. For this simulation, stellar activity jitter is not taken into account.
If 3\,\ms\ jitter is added quadratically, probability is reduced by up to 50\% in
the worst case.}}
    \label{fig:hzplot}
\end{figure}

Fig.~\ref{fig:hzplot} shows the probability of planet detection in the CARMENES
survey as a function of stellar mass and planet semi-major axis {only taking
into account the measured radial velocity uncertainty}. The habitable
zone limits coming from the calibrations given by \citep{Kopparapu14} for
1~$M_{\oplus}$ planets are also plotted. This figure illustrates that, as
expected, the probability of detection is higher for close-in planets, but
still above $\sim$25\% for habitable zone planets around late-type M dwarfs.
{As expected, if we consider also the 3\,\ms\ stellar activity jitter, the
detection probability is reduced (up to 50\% in the worst cases). However,
simultaneous observations using both
CARMENES channels may help to correct this effect.}
This plot is a combination of the planet detectability using periodogram
analyses and our simulated planet rates. The noisy distribution above
$\sim$0.1\,au for low-mass stars is due to small-number statistics, since the
few giant planets generated by our simulations are well above the detection
threshold. Thus, if giant planets are indeed formed around late-type stars
\citep[e.g.,][]{Delfosse98,Johnson07} they can be easily identified, even for
long period orbits.

\section{Conclusions}
\label{sec:conclusions}

In this paper we show that automatic scheduling algorithms are efficient tools
that help to improve the outcome from exoplanet surveys by optimizing the
observation planning and execution. In particular, Artificial
Intelligence techniques are well suited to search for optimal solutions within
the large space of combinations of observations, and can be adapted and
generalized to any kind of survey. For the case of CARMENES we have created a
scheduler, CAST, which takes into account observational constraints and
distributes the telescope time amongst the different targets of the survey. We
demonstrate that with the CARMENES instrument, it is possible to carry out a
spectroscopic survey of a large sample of M-dwarf stars in three years. Our
simulations using a list of 309 M dwarfs show that CAST optimizes the use of
the instrument and can yield over { 60} observations per target, fulfilling all
constraints. Besides, an important advantage of using GAs in the automated
scheduling process is that they guarantee a feasible, consistent, and
near-optimal solution according to the constraints defined in the problem.
CAST can be adapted to the needs of the astronomers in different situations
 during the survey and the night operation. Moreover, due to the fact that it is
 based on a Multi-objective Optimization Problem paradigm by using
Multi-Objective Evolutionary Algorithms, CAST is able to find optimal solutions
with a trade-off between several criteria that can be in conflict with each
other. This aspect provides an important advantage to CAST in order to guarantee
the efficiency of the solutions in terms of use of resources (e.g., telescope
operations) and in terms of science (e.g., planet detection).

In absence of noise sources other than instrumental, our simulations using
recently published planet distribution statistics indicate that these
observations could yield $\sim$118 exoplanets above the CARMENES detection
threshold. {This number should be taken with caution given the still large
uncertainties of planet statistics around M dwarfs, but it could be used a
reference of the planet detection efficiency.}
This means that about { 65\%} of the planets causing radial-velocity
semi-amplitudes larger than 1~\ms\ would be detectable. As expected, the number of
detectable exoplanets lowers with noisier data. Assuming a radial-velocity
jitter due to stellar activity of 3~\ms\ (white noise), the fraction of
detectable planets decreases to { $\sim$16\%} for those with radial-velocity
semi-amplitudes above 1~\ms, resulting in a number around { 28}. However, the
simultaneous observation of radial velocities at optical and near-infrared
wavelengths provides means to counteract this effect and permit the
disentangling of planet radial-velocity signals from stellar activity. In all
cases, most of the potentially detectable planets are super-Earths with masses
below 10~M$_{\oplus}$, and some of them in the habitable zones of their stars.

\section*{Acknowledgement}
We are grateful to the referee for helpful suggestions. 
CARMENES is an instrument for the Centro Astron\'{o}mico Hispano Alem\'{a}n de
Calar Alto in Almer\'{i}a, Spain. CARMENES is funded by the German Max-Planck-Gesellschaft
(MPG), the Spanish Consejo Superior de Investigaciones Cient\'{i}ficas (CSIC), the European Union
through FEDER/ERF funds, and the members of the CARMENES Consortium (Max-Planck-
Institut f\"{u}r Astronomie, Instituto de Astrof\'{i}sica de Andaluc\'{i}a,
Landessternwarte K\"{o}nigstuhl, Institut de Ci\`{e}ncies de l’Espai, Institut
f\"{u}r Astrophysik G\"{o}ttingen, Universidad Complutense de Madrid,
Th\"{u}ringer Landessternwarte Tautenburg, Instituto de Astrof\'{i}sica de Canarias,
Hamburger Sternwarte, Centro de Astrobiolog\'{i}a and the Centro Astron\'{o}mico Hispano-Alem\'{a}n),
with additional contributions by the Spanish Ministry of Economy, the state of Baden-W\"{u}rttemberg,
the German Science Foundation (DFG), the Klaus Tschira Stiftung, and by the Junta de Andaluc\'{i}a.
This project has received funding from the European Union's Horizon 2020 research and innovation
programme under grant agreement No 653477.
We also acknowledge financial support from the Spanish Ministry of Economy and Competitiveness (MINECO)
through grants ESP2013-48391-C4-1-R, ESP2014-57495-C2-2-R, AYA2014-54348-C3-1-R and AYA2014-54348-C3-2-R.
AR acknowledges support from the European Research Council under the FP7 Starting Grant agreement
number 279347 and from DFG grant RE 1664/9-1.

%\end{acknowledgements}

%Bibliography
\bibliographystyle{natbib}
%\bibliography{bibliography}
%\begin{thebibliography}{}

%Appendix
\appendix
\section{Hard constraints computation}
\label{app:constraints}
\subsection{Moon influence} \label{ap:moon}
{ CAST introduces the influence of the Moon on the CARMENES survey
observations by means of a merit function depending on the Moon phase and
the target properties. For each potential target observation $o$ starting at
time $o_{\rm startTime}$, the minimum accepted distance to the Moon $r(o)$ is
computed as a function of the Moon phase $\phi(o_{\rm startTime})$ (fraction of
surface illuminated) following the equation }

\begin{tiny}
\begin{align}
\label{eq:Hmd1}
\begin{split}
% &H_{\rm md}\left(o\right) = \begin{cases} \text{Moon not near the object,} & \mbox{if $dm\left(o_{\rm target},o_{\rm startTime}\right) \geq r\left(o\right)$} \\ \text{Moon near the object,} & \mbox{if $dm\left(o\right) < r\left(o\right)$} \end{cases} ,\\
 &r\left(o\right){\rm =}\left(\left(r_{\rm min}-1\right)\cdot \phi\left(o_{\rm startTime}\right)\right)+1\ .
\end{split}
\end{align}
\end{tiny}

\noindent 
{where the minimum distance during full Moon is set to $r_{\rm min}$=20 deg.
For all targets farther from the Moon a hard constraint
function $H_{\rm mi}$ is computed taking into account the Moon phase and the target
magnitude following the equations }

\begin{tiny}
\begin{align}
\label{eq:Hmi1}
\begin{split}
% H_{\rm mi}(o) & =\begin{cases} \frac{v_{\rm 1}(o)}{v_{\rm 2}(o)} & \mbox{if $\frac{v_{\rm 1} (o)}{v_{\rm 2} (o)} \leq \beta$} \\ 0 & \mbox{if $\frac{v_{\rm 1} (o)}{v_{\rm 2} (o)} > \beta$} \end{cases} ,\\
 &H_{\rm mi}(o){\rm =}\frac{v_{\rm 1} (o)}{v_{\rm 2} (o)}\ ,\\
 &v_{\rm 1}(o){\rm =}1 - \frac{m_{\rm Moon} - m(o_{\rm target}) \cdot (\phi(o_{\rm startTime}))^\alpha}{m_{\rm Moon} - m_{\rm min} \cdot (\phi(o_{\rm startTime}))^\alpha}\ ,\\
 &v_{\rm 2}(o){\rm =}1 - \frac{m_{\rm Moon} - m_{\rm max}}{m_{\rm Moon} - m_{\rm min}}\ .
\end{split}
\end{align}
\end{tiny}

\noindent
{where $m_{\rm moon}$ is the magnitude of the Moon (set to $-$12~mag), 
$m_{\rm min}$ and $m_{\rm max}$ are the minimum and maximum magnitudes of the stars in the sample and $\alpha$ is a power scaling factor, so that observations of faint targets close to near full-moon phases have very low priority.
The denominator $v_{\rm 2}(o)$ is a constant scaling factor (depending on the sample)
to normalize the hard constraint function $H_{\rm mi}$ (so that it can be as well used as a
merit function if desired). All targets below a certain $H_{\rm mi} \leq \beta$ threshold
are taken into account in the scheduler. The $\alpha$
and $\beta$ parameters introduced in these equations are calibrated following \citep{Krisciunas1991}
and Calar Alto sky background conditions \citep{Sanchez08} so that targets are always five magnitudes
brigther than the sky background. We find that values of $\alpha$=10 and $\beta$=0.8
satisfy this criterion.}

\subsection{Integration time} \label{ap:integrationtime}

The estimated integration time is computed with Eq.~\ref{eq:intTime},
where $time_{\rm max}$ is the maximum integration time, $t_{\rm 0}$ is the nominal
integration time, $SN$ is the desired signal to noise ratio, $SN_{\rm 0}$ is the
nominal signal to noise ratio, $m_{\rm t}$ is the magnitude of target $t$, and $m_{\rm 0}$
is the nominal magnitude. All values that do not depend on target $t$, can be parametrized as follows: 

\begin{tiny}
\begin{align}
\label{eq:intTime}
\begin{split}
 &it(t){\rm =}\begin{cases}ct(t) & \mbox{if $ct(t) \leq time_{\rm max}$} \\ time_{\rm max} & \mbox{if $ct(t) > time_{\rm max}$} \end{cases}\ ,\\
 &ct(t){\rm =}t_{\rm 0} \cdot \left(\frac{SN}{SN_{\rm 0}}\right)^2 \cdot 10^{\frac{m_{\rm t}-m_{\rm 0}}{2.5}}\ .
\end{split}
\end{align}
\end{tiny}

{ Reference values for this equation were computed from real CARMENES survey
observations reaching SN$\sim$150, from which radial velocities can be derived with
uncertainties at the level of 1~\ms\, \citep{Reiners10}. The exposure time
calibration as a function of $J$-band magnitude yields  $SN_{\rm 0}$=150 for a
$m_{\rm 0}$=8\,mag star ($J$-band) with and exposure time $t_{\rm 0}$=875\,s.
The maximum exposure time set for any target is $time_{\rm max}$=30\,minutes to
avoid biasing the barycentric correction. }

\subsection{Pointing} \label{ap:point}
The dome of the 3.5m telescope at the Calar Alto Observatory has a
``segmented'' hatch. These segments allow five open window configurations with
the following apertures:
  \begin{itemize}
  \item Window 1: 9\,deg--31\,deg.
  \item Window 2: 26\,deg--46\,deg.
  \item Window 3: 42\,deg--62\,deg.
  \item Window 4: 58\,deg--78\,deg.
  \item Window 5: 73\,deg--92\,deg.
  \end{itemize}
The scheduling algorithm takes into account this configuration of dome
apertures and makes sure that the window is not obscuring the aperture during
the integration. This is done by computing in advance the path of the target
on the sky and the tracking of the dome during the observation. If the hatch
must be moved for a particular target, such target is moved to another time
slot and a different target is chosen.

\subsection{Overhead time} \label{ap:overheadTime}

Eq.~(\ref{eq:Hnot}) describes how to compute the prediction of the
time gap ($H_{\rm not}$) between the end of an observation and the
beginning of a new one, where
         $o_{\rm 1}$ is the completed observation,
         $o_{\rm 2}$ is the next observation,
         $ost$ is the overhead slew time for stabilization defined by the instrument properties,
         $rot$ is the read-out time of the observation,
         and $st$ is the slew time of the telescope that is computed by adding the
         time needed to move the dome ($dt$), the hatch ($ht$) and the telescope ($tt$) from the target of $o_{\rm 1}$ to the target of $o_{\rm 2}$.

\begin{tiny}
\begin{align}
\label{eq:Hnot}
\begin{split}
 &H_{\rm not} \left(o_{\rm 1},o_{\rm 2}\right){\rm =}\begin{cases} st\left(o_{\rm 1},o_{\rm 2}\right), & \mbox{if $st\left(o_{\rm 1},o_{\rm 2}\right) \geq rot\left(o_{\rm 1}\right)$} \\ rot\left(o_{\rm 1}\right), & \mbox{if $st\left(o_{\rm 1},o_{\rm 2}\right) < rot\left(o_{\rm 1}\right)$} \end{cases}\ ,\\
 &st\left(o_{\rm 1},o_{\rm 2}\right){\rm =}ost + dt\left(o_{\rm 1},o_{\rm 2}\right) + ht\left(o_{\rm 1},o_{\rm 2}\right) + tt\left(o_{\rm 1},o_{\rm 2}\right)\ .
\end{split}
\end{align}
\end{tiny}

According to instrument and observatory specifications we assume in our
calculations a telescope and dome slew rate of 1 degree per second, an
stabilization overhead time $ost$=120~s, and a read out time $rot$=40~s.
Finally, the dome hatch takes 60~s to change its position.

\section{Multi-Objective Evolutionary Algorithm design} \label{ap:moea}

\begin{figure}[!t]
\begin{center}
 \includegraphics[width=0.9\columnwidth]{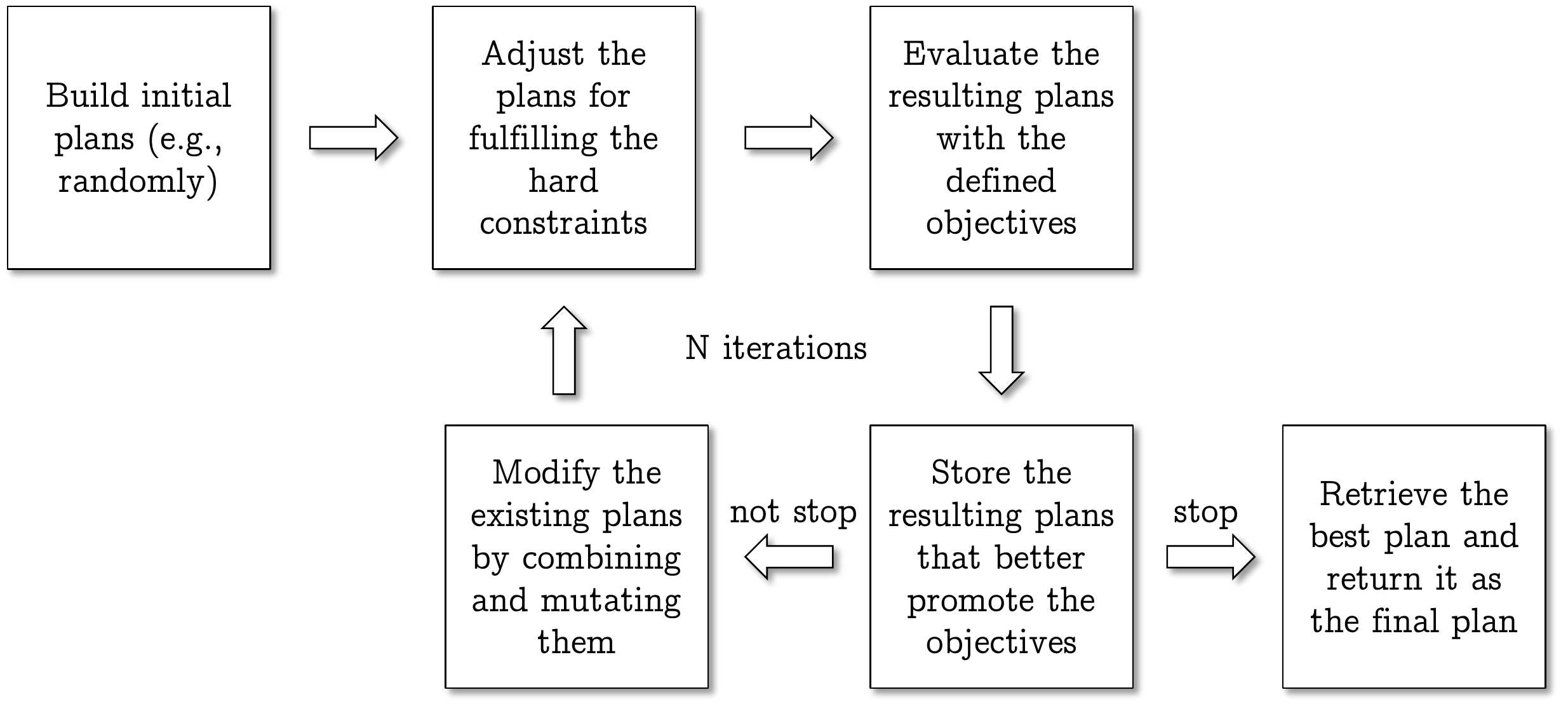}
\end{center}
\vspace*{-0.4cm}
\caption{Cycle followed by a GA.}
\label{fig:GA}
\end{figure}

A MOEA has four main parts: the individual representation, the genetic operators, the objective functions and the selection of the most suitable solution.
As a GA, it follows the process depicted in Fig.~\ref{fig:GA}.

\subsection{Individual representation}
In a Pittsburgh-style GA, each potential solution in the genetic process is
referred to as an individual ($I$) and its representation is based on the definition
of a genotype, which is a set of genes that can have different values, named
alleles \citep{Holland75b, BacarditPhd}. The individual genotype depends on the
type of problem to solve. The first step of the algorithm randomly builds $N_I$
individuals to be assigned to the population.

In NSGA-II \citep{Deb2002}, the individuals in the population are sorted in several fronts
based on non-domination. The first front includes a set of non-dominated individuals
according to the current population, the second front includes the individuals that
are only dominated by the individuals in the first front, and so it continues until
the last of the fronts. The individuals have a rank assigned according to the front
to which they belong. Thus, individuals in the first front have a rank value of
1, individuals in the second one have a rank value of 2, and so on. For computing
if an individual is dominated or non-dominated it is necessary to assign them a
value for each one of the objectives evaluated in the algorithm.

\subsection{Genetic operators} \label{ap:genops}
The GA process is roughly based on applying selection, reproduction (crossover),
mutation, and replacement \citep{Goldberg89, Freitas02} operators for several
iterations, which are described as follows:
%A tournament selection strategy, with a selection probability $p_s$, has been chosen for this problem. This is because it is one of the most widely used selection strategies in GAs and it works efficiently for a wide range of problems \citep{Freitas02}. Our approach has been designed to use elitism in the population, which can contain a maximum of $N_P$ individuals. The crossover and mutation operators are described as follows:
\begin{itemize}
\item \emph{Selection. } $N_S$ parents are selected from the current population
by using a tournament selection strategy \citep{Freitas02}, with a selection probability $p_s$,
based on the rank and crowding distance. The crowding distance is a measure of
how close an individual is to its neighbors according to the value of the
evaluated objectives \citep{Deb2002}. Therefore, an individual is selected if
the rank is lower than the other ones or if crowding distance is greater than
the other ones. After selecting the parents, the crossover and mutation
operators generate two offspring for each pair of parents, obtaining $N_S$ new
individuals.
\item \emph{Crossover. } A crossover operator builds two new individuals from two
previously selected parents. In our case, the two new individuals are obtained by 
using uniform crossover, which is based on
assigning for each gene of the first child the allele of the same gene of the
first parent or the second parent with a probability of 0.5. The alleles of the
genes of each parent not assigned to the first child are copied in the
corresponding genes of the second child. Parents are crossed with a specific
$p_c$ probability, which means that there are some situations where parents are
not crossed and the two offspring are the parents themselves.
\item \emph{Mutation. } The mutation is applied to each gene of every new
individual with a probability of $p_{\mu}$, which means that some genes are not
mutated. Usually, $p_{\mu}$ is a low value because only few genes have to be
mutated in order to make minor changes to the individual, which is the key of
diversity.
\item \emph{Replacement. } The new individuals obtained after the application
of the crossover and mutation operators are evaluated according to the defined
objectives. Next, they are merged with the individuals of the current
population in a temporal one. The temporal population is sorted in non-dominated
fronts. Next, only the best $N_P$ individuals are added to the population used
in the next generation. Individuals are selected based
on their rank and on their crowding distance if they belong to the same front.
\end{itemize}

\subsection{Objective functions}
The definition of these figures of merit (also called fitness functions) is
related to the goals of the problem (i.e., the optimization to be achieved).
Moreover, in order to obtain the results in a reasonable time, these functions
must imply low computational cost because they will be evaluated
many times during the genetic process.

\subsection{Selection of the most suitable solution}
When the last generation of the NSGA-II has been achieved, the algorithm
returns the optimal front of the population (i.e., all the individuals that
belong to the first front, which are non-dominated by any individual of the
rest of the population). This means that, considering that the algorithm has
converged, there is no solution that can improve any objective without
penalizing the other ones. Thus, all individuals have a different trade-off
between objectives but there is no individual better than the other ones.
For this reason, it is necessary to add a new step at the end of the algorithm
to identify the most suitable solution according to some specific criteria.

\end{document}